\documentclass{biometrika}
\makeatletter
\def\ps@plain{\firstpage{\thepage}\let\@mkboth\@gobbletwo\def\@oddhead{}\def\@evenhead{}\def\@oddfoot{}\def\@evenfoot{}}
\makeatother

\usepackage{amsmath}
\usepackage{newtxtext}
\usepackage[subscriptcorrection]{newtxmath}
\usepackage{graphicx}
\usepackage{booktabs,tabularx,array,makecell}
\usepackage{setspace}
\usepackage{url}
\usepackage{xcolor}
\usepackage{comment}

\makeatletter
\newcommand\subsubsection{\@startsection{subsubsection}{4}{\z@}{-9.7pt \@plus -2pt \@minus -2pt}{2pt}{\normalfont\itshape}}
\makeatother

\usepackage{xr}
\makeatletter
\newcommand*{\addFileDependency}[1]{\typeout{(#1)}\@addtofilelist{#1}\IfFileExists{#1}{}{\typeout{No file #1.}}}
\makeatother

\newcommand{\halpha}{\hat{\alpha}}
\newcommand{\balpha}{\boldsymbol{\alpha}}
\newcommand{\bbeta}{\boldsymbol\beta}

\newcommand{\hbbeta}{\hat{\boldsymbol\beta}}
\newcommand{\tbbeta}{\tilde{\boldsymbol\beta}}
\newcommand{\bgamma}{\boldsymbol\gamma}

\newcommand{\bmu}{\boldsymbol{\mu}}
\newcommand{\hmu}{\hat{\mu}}

\newcommand{\bpi}{\boldsymbol\pi}
\newcommand{\hpi}{\hat\pi}

\newcommand{\E}{\mathbb{E}}

\newcommand{\bW}{\textbf{W}}
\newcommand{\tW}{\tilde{W}}
\newcommand{\tbW}{\tilde{\textbf{W}}}
\newcommand{\bx}{\mathbf x}
\newcommand{\bX}{\mathbf X}

\newcommand{\bY}{\mathbf{Y}}

\newcommand{\ind}{\mathbb I}

\newcommand{\Var}{\textup{Var}}
\newcommand{\var}{\textup{Var}}

\newcommand{\ben}{\begin{enumerate}}
\newcommand{\een}{\end{enumerate}}

\title{Modeling Risks with Masked Causes and Variable Exposure}
\author{WENJING MENG$^{1}$, JEFFREY E. MAX$^{2}$ AND FLORIN VAIDA$^{1,*}$}
\affil{$^{1}$Division of Biostatistics and Bioinformatics,\\ Herbert Wertheim School of Public Health and Human Longevity Science,\\ University of California San Diego, 9500 Gilman Drive, La Jolla, California 92093, U.S.A.
\\ $^{2}$Department of Psychiatry, University of California San Diego,\\ 9500 Gilman Drive, La Jolla, California 92093, U.S.A.
\email{w2meng@ucsd.edu} \email{jmax@health.ucsd.edu} \email{fvaida@health.ucsd.edu}\\
$^{*}$Corresponding author.}
\jyear{2026}
\cyear{2026}

\begin{document}

\label{firstpage}
\maketitle

\begin{abstract}
Cohort studies and surveys often record how much time each subject spent under each of two or several treatment conditions and whether an event occurred, but not the condition under which it occurred. We develop two models for such data, for binary and count outcomes. Both start with events under each treatment condition as independent Poisson counts, and estimate rate ratios between treatment and control conditions.
In the binary model, only an indicator of any event is observed, and the event probability under each condition follows a complementary log--log model with a log-exposure offset; development is done for general link functions, including log and logit. The interpretation of the treatment effect changes accordingly. In the count model, the number of events combined over all treatment conditions is observed; overdispersion is handled by Pearson scaling. Parameters are estimated either by the expectation--maximization algorithm or by direct maximization of the observed-data likelihood with Fisher scoring; standard errors are available via observed or expected information. When total exposure is common to all subjects, a relative slope summarizes the treatment effect and, unlike the rate ratio, does not depend on the assumed total exposure. In simulations, interval coverage was near nominal when enough events occurred during each treatment condition. The models are implemented in the R package \textit{maskcave}.

The motivation and immediate application was the Adolescent Brain Cognitive Development study of 10,948 children. Assuming 16 waking hours per day, the adjusted sport-to-non-sport rate ratio for mild traumatic brain injury was 8.78 (95\% confidence interval 4.47--17.27). We applied the count model to concussions during sports reported for 6,379 children at the Year 5 visit: the concussion rate per hour of football was 2.46 (1.37--4.43) times that of the other reported sports, effects that are statistically significant and clinically relevant.
\end{abstract}


\section{Introduction}
In certain studies, individuals are exposed to different treatment conditions for varying amounts of exposure times.  The event of interest is the immediate consequence of exposure to a treatment condition, but the study does not record which of the two or several treatments caused the event.  Inference, however, is possible in large studies, by ascertaining differences in treatment exposures between individuals, and the associated variation in the risk of events.
We develop a statistical setup that uses reported individual-level exposures to each treatment condition to estimate the risk of event during each treatment condition,  when the cause of the observed event is masked.  The actual cause is treated statistically, explicitly or implicitly, as missing data.

This work is motivated by the Adolescent Brain Cognitive Development (ABCD) study, an ongoing longitudinal cohort study in which 11,878 children aged 9--10 years were enrolled at 21 research sites in the U.S. beginning in 2016 \citep{Barch2018}. The study aims to characterize brain, cognitive, behavioral and health development from late childhood through early adulthood, using biopsychosocial data collected at baseline and at regular follow-up visits. Parents report their children's participation in a range of sports and physical activities (hereafter, sports), together with session frequency and duration. The records include an individual history of occurrence of mild traumatic brain injury (mTBI), our outcome of interest, but not whether the injury occurred during sport or another activity.  Our aim is to evaluate the risk of mTBI associated with sports participation (for individual sports or altogether), during the engagement in the activity.

Youth sports participation has been studied for its physical activity and health benefits (\citealp{Shull2020}; \citealp{Bengtsson2025}; \citealp*{Lee2018}) and for its contribution to pediatric injury \citep{Adirim2003}. Because brain injuries in children can have behavioral and psychiatric consequences, the association between sport exposure and mTBI is of clinical interest \citep{Li2013,Max2021,Max2012}. Our previous work examined the risk of mTBI associated with sports participation in the ABCD study, and found higher risk among children who participated in soccer or climbing, and lower odds among those who participated in yoga \citep{Meng2025Sport,Meng2025Sex,Meng2025OI}. Those analyses treated sports participation as binary exposure; here we use the reported amount of time spent in each sport as a refined measure of exposure, and estimate the instantaneous risk of mTBI associated with the sport.  The analyses use cross-sectional ABCD data from baseline.  Starting with Year 5 visit, the records include counts of concussions occurring during sports activities, without attributing the injury to individual sports.  We use these data, together with the exposure to each of several sports that the participant played, to develop a similar model for count data, and compute relative rates of events for different sports.

Poisson processes provide a starting point, since they account for exposure in a natural way, and allow dependence on covariates as a generalized linear model (GLM, \citeauthor{McCullaghNelder1989}, \citeyear{McCullaghNelder1989}).  When counts of one or more events are grouped together, as in our setup, this censored-count Poisson GLM \citep{TERZA1985} is equivalent to a Bernoulli GLM with complementary log-log link \citep{Piegorsch1992}.
\citet{PrenticeGloeckler1978} arrived to this connection in a grouped-time proportional hazards context. The underlying Poisson counts can be treated as unobserved or missing data, with an EM algorithm facilitating maximum likelihood estimation \citep{Dempster1977EM,McLachlanNgNguyen2022}.
\citet*{McMahanWangTebbs2013} developed an EM algorithm with unobserved Poisson counts in the context of regression for current-status survival data. These methods relate a binary event history to the cumulative event intensity over the period before assessment. In our case, the cumulative event intensity under each treatment condition also depends on the exposure to that treatment. To interpret the estimates as rates under each treatment, the reported exposures must be reasonably accurate.  For the ABCD data, the time spent in sports should measure how participants divided their time over the period covered by the event history.

A key feature of our setup is that the cause of the event is masked.  Related problems of masked causes of failure have been addressed in the survival data literature.
\citet{GoetghebeurRyan1995} use conditional cause probabilities to weight failures with two competing causes, which are masked for a subset of the data. \citet{CraiuDuchesne2004} consider a more general setup with several possible masked causes of events, with information provided by second-stage autopsy data. They, as well as \citet{CraiuReiser2006} and \citet{CraiuLee2005}, treat the masked cause as missing data, with estimators computed using an EM algorithm.  In contrast to the competing risks setup, where the failure has one cause, in the case of binary events or counts with multiple treatment exposures, the events may occur independently within each treatment exposure.
For example, a child may have sustained one injury during sport and another outside sport, with the record indicating only that at least one injury occurred. 

Group testing provides another point of comparison with our models. A pooled test, used to make testing for rare events such as HIV infection more efficient, is positive when at least one individual in the pool is positive \citep{PetitoJewell2016}. \citet{LiEtAl2024} develop a model with the status of individuals within the pool as missing data, and maximum likelihood estimation via the EM algorithm.  

In our model, the unobserved indicators refer to events during each treatment condition within an individual.
We call it a variable exposure model, because it's precisely the variation in exposure 
between individuals and treatment conditions that provides information that allows us to tease out the risk specific to each treatment condition. The  treatment conditions are competing but not mutually exclusive causes of the observed events.  An underlying Poisson model provides a natural justification and it leads to Bernoulli GLMs within each treatment condition, with complementary log-log link.  However, our development is for a general link function, and does not require such an underlying process.  The treatment effect is expressed as the treatment coefficient in the GLMs, with the log-exposures as offsets in the models.  The maximum likelihood estimator is computed via the EM algorithm and via direct Fisher scoring \citep{Lange2013}, with asymptotic variances given by inverse Fisher information and via Louis' formula \citep{Louis1982}.  An extension to multiple treatment conditions is given.
A similar model is developed when the observed outcome is a number of counts of events, representing the individual-level total due to different, masked causes (Section~\ref{sec:counts}). Extensive simulations assess the validity of estimation, including bias, confidence-interval coverage, comparison of computation speed of proposed algorithms, and their behavior from poor starting values. As a case study we relate the occurrence of mTBI to playing sports, using the ABCD study baseline data (binary outcome), and counts of traumatic brain injury and their relation to playing different sports, using Year 5 ABCD study data (count outcome).  Supplementary materials include detailed proofs and derivation, and additional analyses.

\section{Methods}
\label{sec:methods}
\subsection{Statistical Model Setup} \label{sec:setup}

For generic subject $i$, we are interested in the causal effect of a binary exposure $Z_i$ on a binary outcome $Y_i$.  We adjust for relevant confounders, in the form of covariate vector $\mathbf{x}_i=(x_{i1},\ldots,x_{iq})^\top$; $\bx_i$ may be a balancing score, such as a propensity score for treatment $Z_i$ \citep{RosenbaumRubin1983, Imbens2000}, or a generalized propensity score \citep{HiranoImbens2004} or propensity function  \citep{ImaiVanDyk2004}. Each subject is exposed to both treatment and control conditions, for total known exposures $h_{1i}$ and $h_{0i}$ respectively, with $h_i=h_{0i}+h_{1i}$.  Exposure $h_i$ may be a time, count, area, etc. \citep{Diggle2013}.  Some subjects may have a single exposure, $h_{0i}=0$, or $h_{1i}=0$.

When event $Y_i=1$ occurs, its cause is masked: it is unknown if an event happened during treatment exposure, control exposure, or both.  We are interested in  relative and absolute event rates associated with treatment and control, conditional on covariates, as defined below.  

To model $Y_i$ as a function of exposures $h_{0i}, h_{1i}$, assume momentarily that the event is repeatable, with unmeasured event counts $W_{0i}, W_{1i}$ during the control and treatment exposures, assumed independent.  Let $Y_{0i} = \min(W_{0i}, 1), Y_{1i} = \min(W_{1i}, 1)$ be the indicators of any events occurring during each exposure.  Then $Y_i= \min(Y_{0i}+Y_{1i},1)=\min(W_{0i}+W_{1i},1)$, the indicator of at least one event occurring during either exposure.  

The counts $W_{0i}, W_{1i}$ can be modeled via a Poisson log-linear model
\[
 W_{0i}\sim\operatorname{Poisson}(\mu_{0i}),\qquad
 W_{1i}\sim\operatorname{Poisson}(\mu_{1i}),
\]
with means $\mu_{0i} = h_{0i}\lambda_{0i}, \mu_{1i}=h_{1i}\lambda_{1i}$, proportional to exposures $h_{0i}, h_{1i}$.  The rates $\lambda_{0i}, \lambda_{1i}$ are assumed to follow the log-linear model
\[
\log\lambda_{0i}=\alpha_{0}+\bgamma^\top\bx_i, \qquad
\log\lambda_{1i}=\alpha_{0}+\alpha_1+\bgamma^\top\bx_i,
\]
for appropriate parameters $\bbeta = (\alpha_0, \alpha_1, \bgamma)$.
(Our notation does not distinguish between row and column vectors, unless necessary.)  The model above assumes treatment effect homogeneity, i.e., no treatment-by-covariate interaction \citep{ImbensRubin2015}; extensions to heterogeneous treatment effects are possible.
The parameter of primary interest is $\alpha_1$:
$\exp(\alpha_1) = \lambda_{1i}/\lambda_{0i}$ is the conditional rate ratio of events
representing the effect of treatment on covariate-adjusted event rate, per unit of exposure $h_{0i}=h_{1i}=1$ \citep{GreenlandRobinsPearl1999}.
If the covariates $\bx$ are centered, $\exp(\alpha_0), \exp(\alpha_0+\alpha_1)$ are interpretable, as event rates per unit of exposure, under control and treatment conditions, for an individual with $\bx=0$. 

Let $\pi_{ki} = \pi_{ki}(\bbeta) = P(Y_{ki}=1) = P(W_{ki}\geq 1)$, for $k=0,1$.  Then $1-\pi_{1i} = P(W_{1i}=0) = e^{-\mu_{1i}}$, and
$\log(-\log(1-\pi_{1i})) = \log \mu_{1i}$.  
Similarly,
$\log(-\log(1-\pi_{0i})) = \log \mu_{0i}$.  Put $\bx_{0i}=(1,0,\bx_i),
\bx_{1i}=(1,1,\bx_i)$, and let $g(\pi) = \log(-\log(1-\pi))$ be the complementary log-log link.  Therefore, 
\begin{equation} \label{eq:glm}
g(\pi_{0i}) = \log h_{0i} + \bbeta^\top \bx_{0i}
\qquad
g(\pi_{1i}) = \log h_{1i} + \bbeta^\top \bx_{1i}.
\end{equation}
$Y_{0i}, Y_{1i}$, which are censored versions of Poisson variables $W_{0i}, W_{1i}$ \citep{TERZA1985}, follow Bernoulli GLMs with complementary log-log link $g$ \citep{Piegorsch1992} and same linear predictors as $W_{0i}, W_{1i}$, namely
$\eta_{0i} = \log h_{0i} + \bbeta^\top \bx_{0i}$, 
$\eta_{1i} = \log h_{1i} + \bbeta^\top \bx_{1i}$.  These include  offsets $\log h_{0i}, \log h_{1i}$ \citep{McCullaghNelder1989,McMahanWangTebbs2013}. 
For individuals $i=1,\ldots, N$ we measure independent variables $Y_i$, with exposures $h_{1i}, h_{0i}$, and covariates $\bx_i$.
Individual event counts $W_{0i}, W_{1i}$ during $Z_i=0$ and $Z_i=1$ respectively, as well as $Y_{0i} = \min(W_{0i},1)$ and $Y_{1i} = \min(W_{1i},1)$ are missing data.
The GLMs (\ref{eq:glm}) require $h_{1i}, h_{0i} > 0$.  The case $h_{1i}=0$ implies $Z_i=0$ throughout the entire exposure $h_i$; then  $Y_i = Y_{0i}$, and $W_{1i}=Y_{1i}=0$. A similar approach is taken if $h_{0i}=0$.
In these cases, the cause of event is known. 

We develop two methods to derive the maximum likelihood estimator of $\bbeta=(\alpha_0, \alpha_1, \bgamma)$ and its asymptotic variance: the EM algorithm, and a direct Fisher scoring algorithm.  While complementary log-log link has best conceptual justification, the model can be developed for a generic link function $g$ starting with $Y_{0i}, Y_{1i}$ directly, without the use of Poisson counts $W_{0i}, W_{1i}$.  The choice of link functions is discussed in Section~\ref{sec:extensions}.

\subsection{Maximum likelihood estimation via the EM algorithm}
\label{sec:em}

Let the observed data vector be $\bY = (Y_1, \ldots, Y_N)^{\top}$.  For simplicity, the dependence of the quantities below on exposures and on covariate matrix  
$\bX$ with rows $\bx_1, \ldots, \bx_N$ is suppressed in notation.
The maximum likelihood estimator (MLE) maximizes the Bernoulli log-likelihood 
\begin{equation} \label{eq:loglik}
\ell(\bbeta;\bY) = \sum_{i=1}^N \{Y_i\log \pi_i+ (1-Y_i)\log(1-\pi_i)\},
\end{equation}
where $\pi_i = \pi_i(\bbeta) = P(Y_i=1) = 1-(1-\pi_{0i})(1-\pi_{1i}), i=1,\ldots, N$.  Given independent missing data vectors $\bY_0=(Y_{01}, \ldots, Y_{0N})^{\top}, \bY_1=(Y_{11}, \ldots, Y_{1N})^{\top}$, the complete-data log-likelihood is the sum of log-likelihoods for $\bY_0$ and $\bY_1$, $\ell_0(\bbeta;\bY_0)$ and $\ell_1(\bbeta;\bY_1)$:
\begin{eqnarray}
 \ell_c(\bbeta;\bY_0, \bY_1)
 &=&\ell_0(\bbeta;\bY_0) + \ell_1(\bbeta;\bY_1) \nonumber
 \\
 &=&\sum_{k,i}
 \{Y_{ki}\log\pi_{ki}+(1-Y_{ki})\log(1-\pi_{ki})\}.
 \label{eq:complete-loglik}
\end{eqnarray}
The sum $\sum_{k,i}$ is extended over $k=0,1$ and $i=1,\ldots, N$, with the convention $0\cdot \log 0$ corresponding to cases $h_{ki}=0$.

The complete-data score and expected information are
\begin{eqnarray}
    s_c(\bbeta;\bY_0, \bY_1) &=& \frac{\partial \ell_c(\bbeta)}{\partial \bbeta}
    = \sum_{k,i}(Y_{ki}-\pi_{ki})w_{ki}\bx_{ki}
    \label{eq:complete-score}
    \\
    I_c(\bbeta) &=& \sum_{k,i} v(\pi_{ki}) w_{ki}^2 \bx_{ki}\bx^\top_{ki},
    \label{eq:complete-info}
\end{eqnarray}
where $v(\pi)=\pi(1-\pi)$, $w_{ki}=\frac{G'(\eta_{ki})}{v(\pi_{ki})}$,  and $G=g^{-1}$ is the inverse link.  

Following \citet{Dempster1977EM}, define the EM function
$
Q(\bbeta|\tbbeta) = \E\{\ell_c(\bbeta)\mid\bY,\tbbeta\}.
$
The EM algorithm determines the MLE $\hbbeta$ as the limit  of the sequence 
$\bbeta^{(1)}, \bbeta^{(2)}, \ldots$, where at step $t$, $\bbeta^{(t+1)}$ is obtained as
$ \bbeta^{(t+1)} = \arg\max_{\bbeta} Q(\bbeta|\bbeta^{(t)}). 
$
Let $\tbbeta=\bbeta^{(t)}$.  The E-step of the algorithm involves the computation of $Q(\bbeta|\tbbeta)$:
\begin{equation}
 Q(\bbeta|\tbbeta) = \E\{\ell_c(\bbeta;\bY_0,\bY_1)\mid\bY,\tbbeta\}
 = \ell_c(\bbeta;\tilde{\bY}_0, \tilde{\bY}_1)
 \label{eq:qfunction}
\end{equation}
where $\tilde{\bY}_0 = \E(\bY_0|\bY, \tbbeta), \tilde{\bY}_1=\E(\bY_1|\bY,\tbbeta)$.  Specifically, for $k=0,1$, $\tilde{\bY}_k$ has elements 
$$\tilde{Y}_{ki} = \frac{\pi_{ki}(\tbbeta)}{\pi_i(\tbbeta)} \,Y_i.$$
Indeed, for $k=0,1$, if $Y_i=0$, then $Y_{ki}=0$ and $\E(Y_{ki}|Y_i,\tbbeta)=0$.  If $Y_i=1$, $\E(Y_{ki}|Y_i,\tbbeta) = P(Y_{ki}=1, Y_{i}=1|\tbbeta)/P(Y_{i}=1|\tbbeta) = \pi_{ki}(\tbbeta)/\pi_i(\tbbeta)$.  

$Q(\bbeta|\tbbeta)$ has the form of a quasi-Bernoulli log-likelihood in $\bbeta$ for fractional responses $\tilde{Y}_{ki}$ with linear predictors $\eta_{ki}$, link function $g$, and variance function $v(\pi)=\pi(1-\pi)$\citep{Wedderburn1974,PapkeWooldridge1996}.  It can be maximized via the usual Fisher scoring algorithm, using update steps
$ \bbeta \to \bbeta + I_Q(\bbeta)^{-1} s_Q(\bbeta)$, where $s_Q(\bbeta), I_Q(\bbeta)$ are the quasi-score and expected quasi-information functions given by
\begin{eqnarray*}
s_Q(\bbeta) &=& \frac{\partial Q(\bbeta|\tbbeta)}{\partial \bbeta}
= s_c(\bbeta;\tilde{\bY}_0, \tilde{\bY}_1)
= \sum_{k, i} (\tilde{Y}_{ki}-\pi_{ki}) w_{ki} \bx_{ki}
\end{eqnarray*}
and $I_Q(\bbeta) = I_c(\bbeta)$.
Since the M-step involves iteration nested within the EM algorithm sequence, it does not need to be run to convergence: one or several updates that increase Q are sufficient, if it leads to overall speed gains. This is the EM gradient (EMG) algorithm of \citet{Lange1995EMGradient}, a special case of the generalized EM \citep{Dempster1977EM}, that preserves the speed of convergence of the full M-step algorithm.   

Supplementary Section~\ref{app:mstep-general-links} gives the score, Hessian and expected information of the M-step for the complementary log--log, log and logit links.

At each step, the EM algorithm climbs up the log-likelihood function, and its convergence to a stationary point is guaranteed \citep{Wu1983,Vaida2005}.  

Upon convergence to $\hbbeta$, the asymptotic variance of $\hbbeta$ is estimated by $I(\hbbeta)^{-1}$.  This is computed as follows:
Let $u_{ki} = w_{ki}\pi_{ki}/\pi_i$, and $\bx_{Bi} = u_{0i}\bx_{0i} + u_{1i} \bx_{1i}$, $i=1,\ldots, N$.
The observed data expected information can be evaluated via
\begin{equation} \label{eq:info}
I(\bbeta) =
\var\{\E(s_c(\bbeta;\bY_0,\bY_1)\mid\bY)\} = 
\sum_{i=1}^N v(\pi_i) \bx_{Bi}\bx_{Bi}^\top,
\end{equation}
or $I(\bbeta)=\bX_B^\top D_B\bX_B$,
where $\bX_B$ is the matrix with rows $\bx_{B1}, \ldots, \bx_{BN}$, and $D_B$ is diagonal, with diagonal elements $v(\pi_1), \ldots, v(\pi_N)$.

A line of research, based on ancillary arguments, proposes the use of inverse observed information, $J(\hbbeta)^{-1}$, as a better alternative to $I(\hbbeta)^{-1}$ for estimating $\var(\hbbeta)$ \citep{Cox1958,EfronHinkley1978,Efron1975,LindsayLi1997}. 
The observed information 
can be computed using Louis' formula \citep{Louis1982},
\begin{equation} \label{eq:obs_info}
J(\hbbeta) = \E\left[J_c(\hbbeta;\bY_0,\bY_1)\mid \bY, \hbbeta\right] -
\E[s_c(\hbbeta;\bY_0,\bY_1) s_c(\hbbeta;\bY_0,\bY_1)^\top|\bY,\hbbeta],
\end{equation}
where $J_c(\bbeta;\bY_0,\bY_1)$ is the observed information of complete data $(\bY_0,\bY_1)$.
Proof for (\ref{eq:info}) and a formula for (\ref{eq:obs_info}) appear in Supplementary Section~\ref{app:louis}.  In Supplementary Table~\ref{tab:information_comparison} we show results of simulations comparing the performance of the two variance estimators, $I(\hbbeta)^{-1}$ and $J(\hbbeta)^{-1}$.

As an added benefit, the EM algorithm provides direct estimates of probabilities of event $Y_i=1$ having occurred due to treatment or control exposures.  Specifically, for $Y_{i}=1$, the probabilities that for subject $i$ an event occurred during treatment exposure, during control exposure, and during both treatment and control exposure are $\frac{\hpi_{1i}}{\hpi_i}$, $\frac{\hpi_{0i}}{\hpi_i}$, and $\frac{\hpi_{0i}+\hpi_{1i}-\hpi_i}{\hpi_i} = \frac{\hpi_{0i} \hpi_{1i}}{\hpi_i}$, where $\hpi_{0i}=\pi_{0i}(\hbbeta), \hpi_{1i}=\pi_{1i}(\hbbeta), \hpi_i=\pi_i(\hbbeta)$.
Summing these probabilities over subjects gives the expected numbers of subjects with at least one event during each exposure, $\hat{N}_k=\sum_{i=1}^N Y_i\hpi_{ki}/\hpi_i$, $k=0,1$. 

\subsection{Maximum likelihood estimation via direct optimization}
\label{sec:marquardt}

The MLE $\hbbeta$ can be computed by direct maximization of the likelihood. Since the log-likelihood (\ref{eq:loglik}) is in standard Bernoulli form, modified forms of GLM formulas apply for the score and information functions, and the MLE can be obtained via Fisher scoring.
 Standard derivations show that the score and expected information are
\begin{eqnarray} \label{eq:score_direct}
s(\bbeta) &=& \sum_{i=1}^N(Y_i-\pi_i)\bx_{Bi}= \bX_B^\top (\bY-\bpi)
\\
I(\bbeta) &=& \sum_{i=1}^N v(\pi_i) \bx_{Bi}\bx_{Bi}^\top = \bX_B^\top D_B\bX_B,
\label{eq:info_direct}
\end{eqnarray} 
with (\ref{eq:info_direct}) consistent with (\ref{eq:info}). Supplementary Section~\ref{app:direct-likelihood} gives general proofs, and formulas for special cases of complementary log--log, log, and logit links (Supplementary Tables~\ref{tab:link-ef} and~\ref{tab:link-weights}).

A Fisher scoring algorithm updates the parameter sequence via the step $\bbeta \to \bbeta + I(\bbeta)^{-1} s(\bbeta)$, until convergence to $\hbbeta$.  
Alternatively, a Newton-Raphson algorithm can be used, with update step
$\bbeta \to \bbeta + J(\bbeta)^{-1} s(\bbeta)$.  
The MLE variance is estimated as
$\widehat{\Var}(\hbbeta) = I(\hbbeta)^{-1}$ and
$\widehat{\Var}(\hbbeta) = J(\hbbeta)^{-1}$, respectively.
Neither method is guaranteed to increase the log-likelihood at every step, and it may fail to converge from poor starting values. A step-control version $\bbeta\to\bbeta+\epsilon I(\bbeta)^{-1}s(\bbeta)$, with $\epsilon$ set initially to 1 and halved until the log-likelihood increases (step halving), prevents a decrease at any iteration \citep{Marschner2011}. In contrast, an EM iteration never decreases the log-likelihood, and under regularity conditions it converges to a stationary point of the likelihood \citep{Wu1983, Vaida2005}, although convergence can be slow \citep{JamshidianJennrich1997}. A practical strategy is therefore to start with a number of EM iterations and then switch to Fisher scoring or Newton--Raphson, as in some linear mixed-effects models software \citep{BatesPinheiro1998,PinheiroBates2000}. Section~\ref{sec:simulation_computation} compares EM,  EMG, and direct Fisher scoring in the simulations.

\subsection{Model extensions and additional considerations}
\label{sec:extensions}

\subsubsection{Choice of link function}
The Poisson counts $W_{0i}, W_{1i}$ introduced in Section~\ref{sec:methods}, combined with the complementary log-log link $g$, allow a natural interpretation of exposures $h_{0i}, h_{1i}$, and of the treatment effect $\exp(\alpha_1)$, as an underlying conditional rate ratio of events.  However, these assumptions are not necessary: model specification simply requires independent Bernoulli GLMs for $Y_{0i}$ and $Y_{1i}$ with common link $g$ and linear predictors $\eta_{0i}=\log h_{0i}+\bbeta^\top \bx_{0i}, \eta_{1i}=\log h_{1i} + \bbeta^\top \bx_{1i}$, and $Y_i = \min(Y_{0i}+Y_{1i}, 1)$.  Other link functions are possible, such as log, $g(\pi)=\log\pi$, or logit, $g(\pi) = \log\left(\frac{\pi}{1-\pi}\right)$.  
The log and logit links can be justified as approximations to the complementary log-log link for small expected means $\pi_{ki}\approx 0$ \citep{McCullaghNelder1989}.  The exposure $h_{ki}$ is typically incorporated via a $\log h_{ki}$ offset under these links as well \citep{Wacholder1986,ZhangYu1998}. 
For unit exposures $h_{0i}=h_{1i}=1$ and covariates $\bx_i$, we have 
$$
\exp(\alpha_1)
= \frac{\exp(g(P(Y_{1i}=0|\bx_i, h_{1i}=1))}{\exp(g(P(Y_{0i}=1|\bx_i, h_{0i}=1))}.
$$  For log link, treatment effect $\exp(\alpha_1)$ is the conditional risk ratio, $\frac{P(Y_{1i}=1|\bx_i, h_{1i}=1)}{P(Y_{0i}=1|\bx_i, h_{0i}=1)}$.  For logit link, $\exp(\alpha_1)$ is the conditional odds ratio.  

Simulation results and a case study presented in Sections~\ref{sec:simul} and~\ref{sec:application_primary} compare results using complementary log-log, log, and logit links.

\subsubsection{Average treatment effect among the treated}
Conditional treatment effects given by model (\ref{eq:loglik}) can be converted to population-level (marginal) average treatment effects, using  causal inference methodology \citep{HernanRobins2020,DegtiarRose2023,MullerMacLehose2014,DanielZhangFarewell2021}. 
However, modifications are needed, taking into account exposures $h_{0i}, h_{1i}$. 
The causal estimand of interest in this case is $\E(Y_{1i}|h_{0i}=h)-\E(Y_{0i}|h_{0i}=h)$, the difference in expected values of event $Y_i$ under treatment and control conditions, for same exposure $h$.  The excess number of events in sample due to treatment can be estimated by subtracting, from the expected number of events under observed exposures $h_{1i}$, the expected number of events when   all exposure is converted to control condition, i.e. with treatment exposure $0$ and control exposure $h_{0i}+h_{1i}$; standard errors and confidence intervals may be assessed via bootstrap \citep{SnowdenRoseMortimer2011} or the delta method \citep{GraubardKorn1999}.
An in-depth treatment of this issue will be pursued in a separate report.

\subsubsection{Common total exposure}
In our motivating example, the total exposure $h_i$ represents the total time of exposure to mTBI for child $i$.  We chose this, somewhat arbitrarily, to be the common wake time, i.e., $h_i=h=16$ hours/day for all children.  Given sports exposures $h_{1i}$, changing the choice of $h$ (e.g., $h=12$ or $h=24$ hours/day) amounts to a simple reparametrization of the log-likelihood function, that changes the treatment effect $\exp(\alpha_1)$ in a deterministic way.  The data does not help in choosing $h$, besides the trivial constraint $h \geq \max_{i=1}^N h_{1i}$.
The following result holds:

\begin{theorem}\label{thm:common-exposure}
For all $i=1,\ldots, N$, assume treatment exposures $h_{1i}$ are given,  common exposure $h_i =h$, $h\geq h_{1i}$, and control exposures are derived as $h_{0i}=h-h_{1i}$. 
For complementary log-log link, $\rho=(e^{\alpha_1}-1)/h$ is invariant to the choice of $h$.  That is, log-likelihood (\ref{eq:loglik}) depends on $\alpha_0, \alpha_1$, and $h$ only through $(\log h + \alpha_0, \rho)$. 
\end{theorem}

\begin{proof}
Under the complementary log-log link model, with linear predictors $\eta_{0i}=\log h_{0i}+\alpha_0 + \bgamma^\top \bx_i$, 
$\eta_{1i}=\log h_{1i}+\alpha_0 + \alpha_1 + \bgamma^\top \bx_i$, we have that
\begin{eqnarray*}
\log(-\log(1-\pi_i)) &=& \log(-\log(1-\pi_{0i})-\log(1-\pi_{1i}))
\\&=& \log(\exp(\eta_{0i})+\exp(\eta_{1i}))
\\&=& \alpha_{0}+\bgamma^\top \bx_i+\log(h_{0i}+h_{1i}e^{\alpha_1})
\\&=& (\log h +\alpha_0) + \bgamma^\top \bx_i+\log(1+h_{1i}\rho)
\end{eqnarray*}
where $\rho = (e^{\alpha_1}-1)/h$.  Log-likelihood (\ref{eq:loglik}) is a function of $\pi_1,\ldots, \pi_N$, and depends on $h, \alpha_0, \alpha_1$ only through $(\log h+\alpha_0, \rho)$.  
\end{proof}

Given treatment exposures $h_{1i}$, $\rho$ is invariant to the choice of $h$.  However, the treatment effect $\exp(\alpha_1) = 1+h\rho$ depends linearly on $h$.  

The invariant $\rho$ can be interpreted in connection to the Poisson model.  Since
$$\rho = (e^{\alpha_1}-1)/h = \frac{\lambda_{1i}-\lambda_{0i}}{h\lambda_{0i}},$$
$\rho$ is the difference in expected event counts under treatment and control conditions, per unit of exposure, relative to  the expected counts over entire exposure $h$ under control condition.

Put $\Lambda(0)=h\lambda_{0i}$.  The expected total event count for subject $i$ is 
$\Lambda(h_{1i})=e^{\eta_{0i}}+e^{\eta_{1i}} = \Lambda(0)(1+h_{1i}\rho)$. 
This is the form of the linear relative-risk and relative-rate models of \cite{PrenticeMason1986} and \cite{Richardson2008}.

Invariants such as $\rho$ do not exist for other common link functions $g$, such as the log or logit. Supplementary Section~\ref{app:waking} gives the derivations and conditions for comparing the event probabilities $\pi_i$ as functions of $h_{1i}$, at fixed covariate values and different values of $h$.

\subsubsection{Multiple treatment conditions}

So far we considered a binary option for the treatment variable $Z$, $Z=0$ or $1$.  More generally, $K$ treatment options may be available, corresponding to $Z=0, 1, \ldots, K-1$, with $Z=0$ as the reference.  The model above extends directly, with $h_{0i}, \ldots, h_{K-1,i}$ being total known exposure measures for treatments $Z=0, Z=1, \ldots, Z=K-1$, with corresponding independent Bernoulli latent outcomes $Y_{ki}$, with means $\pi_{ki}$, given by $g(\pi_{ki})=\eta_{ki}=\log h_{ki} + \alpha_0+\alpha_{k}+\gamma'\bx_{i}$, $k=0, \ldots, K-1, i=1,\ldots, N$.  The parameter is $\bbeta = (\balpha, \bgamma)$, with $\balpha=(\alpha_0, \alpha_1,\ldots, \alpha_{K-1})$, with rate ratios $\exp(\alpha_k)$ representing conditional, individual-level effects of treatment $k=1,\ldots, K-1$ versus the reference $Z = 0$.  Formulas (\ref{eq:loglik})-(\ref{eq:info_direct}) extend naturally to this case, see Supplementary Section~\ref{app:proofs-methods-n}.

\section{A model for counts}
\label{sec:counts}

In the same setting and notation of Section~\ref{sec:methods}, assume that, for subjects $i=1,\ldots, N$, $W_{0i}$ and $W_{1i}$ are independent Poisson counts of events under treatment and control conditions, with means $\mu_{ki}=g^{-1}(\eta_{ki})$, where $\eta_{ki} = \log h_{ki} + \bbeta^\top \bx_{ki}$, $k=0,1$.   The link function $g$ is usually the natural logarithm, $g(\mu) = \log(\mu)$, with inverse $G(\eta) = \exp(\eta)$.  Let $W_i = W_{0i}+W_{1i}$, the total number of events for subject $i$.  Assume now that, instead of observing the occurrence of binary outcome $Y_i = \min(W_i, 1)$ we observe the total count $W_i$, but individual counts $W_{0i}, W_{1i}$ are missing data.   We are interested in inference regarding parameter $\bbeta=(\alpha_0,\alpha_1,\bgamma)$, with primary interest residing in conditional rate ratio $\exp(\alpha_1)$.  For logarithmic link,  mean $\mu_{ki}$ is proportional to exposure $h_{ki}$, $\exp(\bbeta'\bx_{ki})$ is the individual-level event rate per unit of exposure, and $\exp(\alpha_1)$ is the conditional rate ratio of events, under the treatment versus the control conditions.  As in the binary case, population-level treatment effects require careful construction, taking into account the joint distributions of log-exposures and covariates $(\log h_{0i}, \log h_{1i}, \bx_i)$.

For each subject $i=1,\ldots,N$, suppose that we observe the total
event count $W_i=W_{0i}+W_{1i}$, exposures $h_{0i},h_{1i}$, and
covariates $\bx_i$. The event counts during control and treatment
exposure, $W_{0i}$ and $W_{1i}$, are missing data. 
Independence of $W_{0i}, W_{1i}$ implies
$W_i\sim\operatorname{Poisson}(\mu_i)$ and
$(W_{ki}|W_i) \sim \textup{Binomial}(W_i, \mu_{ki}/\mu_i)$,
where
$\mu_i=\mu_{0i}+\mu_{1i}=G(\eta_{0i})+G(\eta_{1i})$.  (If $h_{ki}=0$, by convention $W_{ki}=0$ with probability 1, and $\mu_{ki}=0$.)  We estimate $\bbeta$ via the MLE $\hbbeta$, the maximizer of log-likelihood
\begin{equation}
 \ell(\bbeta;\bW)
 =\sum_{i=1}^N\{W_i\log\mu_i-\mu_i\},
 \label{eq:counts-loglik}
\end{equation}
where $\bW=(W_1,\ldots,W_N)^\top$, and terms unrelated to $\bbeta$ are omitted.

\subsection{Maximum likelihood estimation via the EM algorithm}
\label{sec:counts-em}

Let $\bW_k=(W_{k1}, \ldots, W_{kN})$, $k=0,1$, and $\bW=(W_1,\ldots, W_N)$. 
Furthermore, denote $\mu_{ki}=\mu_{ki}(\bbeta), \mu_i=\mu_i(\bbeta)$, for all $k,i$, to mark their dependence on $\bbeta$. 
For complete data $(\bW_0,\bW_1)$, the log-likelihood is
\begin{equation*}
\ell_c(\bbeta;\bW_0, \bW_1) = \sum_{k,i}(W_{ki}\log\mu_{ki}-\mu_{ki})
\end{equation*}
where sums extend over all $k=0,1$ and $i=1,\ldots,N$, with the convention $0\cdot \log 0 = 0$, corresponding to cases $h_{ki}=0$.  The complete-data score and expected information are
\begin{eqnarray}
    s_c(\bbeta;\bW_0, \bW_1) &=& \frac{\partial \ell_c(\bbeta)}{\partial \bbeta}
    = \sum_{k,i}(W_{ki}-\mu_{ki})w_{ki}\bx_{ki}
    \label{eq:complete-score-count}
    \\
    I_c(\bbeta) &=& \sum_{k,i} v(\mu_{ki}) w_{ki}^2 \bx_{ki}\bx^\top_{ki},
    \label{eq:complete-info-count}
\end{eqnarray}
where $v(\mu)=\mu$ is the Poisson variance function, and
$w_{ki}=\frac{G'(\eta_{ki})}{v(\mu_{ki})}$.  Due to GLM properties, these formulas map perfectly on those for binary data in (\ref{eq:complete-score}, \ref{eq:complete-info}).
For log link, $G'(\eta_{ki}) = v(\mu_{ki}) = \mu_{ki}$, and $w_{ki}=1$.

The MLE $\hbbeta$ is the limit of the sequence $\bbeta^{(1)},\bbeta^{(2)},\ldots$, defined iteratively as 
$\bbeta^{(t+1)} = \arg\max_{\bbeta} Q(\bbeta|\bbeta^{(t)})$, where
\begin{equation} \label{eq:q-count}
    Q(\bbeta|\tbbeta) = \E[\ell_c(\bbeta;\bW_0,\bW_1)|\bW, \tbbeta]
    = \ell_c(\bbeta;\tbW_0, \tbW_1),
\end{equation}
where for all $k,i$, $\tW_{ki} = \E(W_{ki}|W_i,\tbbeta) = \frac{\mu_{ki}(\tbbeta)}{\mu_i(\tbbeta)}W_{i}$, and $\tbW_k=(\tbW_{k1},\ldots, \tbW_{kN}), k=0,1$.
Given $\tbbeta = \bbeta^{(t)}$, the E-step computes $\tbW_0, \tbW_1$;
the M-step determines $\bbeta^{(t+1)}$ as the limit of the Fisher scoring sequence $\bbeta \to \bbeta +I_c(\bbeta)^{-1}\cdot s_c(\bbeta;\tbW_0,\tbW_1)$, with $s_c, I_c$  given by (\ref{eq:complete-score-count}, \ref{eq:complete-info-count}).  
Similarly to the binary case, at E-step $W_{ki}$ is replaced in complete-data log-likelihood by its conditional expectation, which is a fraction of the observed data.
The M-step approach is justified by the fact that the expected log-likelihood in (\ref{eq:q-count}) is formally identical to a quasi-likelihood with identity variance function, typically maximized via Fisher scoring \citep{Wedderburn1974,McCullaghNelder1989}
Upon convergence, the variance of the MLE is estimated as $\widehat{\Var(\hbbeta)} = I(\hbbeta)^{-1}$, where $I(\hbbeta)$ is the expected information of $\bW$.  This computation is analogous to the binary case: for $i=1,\ldots, N$, let $u_{ki} = w_{ki}\mu_{ki}/\mu_i$, and put $\bx_{Ci} = u_{0i}\bx_{0i} + u_{1i}\bx_{1i}$.  Let $\bX_C$ be the matrix with rows $\bx_{C1},\ldots, \bx_{CN}$, and $D_C$ be the diagonal matrix with diagonal elements $\mu_1, \ldots, \mu_N$.  We show in Supplementary Section~\ref{app:count-info-estep} that
\begin{equation} \label{eq:info-count}
    I(\bbeta) = \bX_C^\top D_C \bX_C.
\end{equation}
Supplementary Section~\ref{app:count-info-louis} derives $J(\hbbeta)^{-1}$ as an alternative estimator of $\Var(\hbbeta)$, via Louis' formula.

\subsection{Direct likelihood maximization}
\label{sec:counts-direct}

The Poisson log-likelihood (\ref{eq:counts-loglik})  may also be maximized directly, using a Fisher scoring algorithm with transitions $\bbeta \to \bbeta + I(\bbeta)^{-1}\cdot s(\bbeta;\bW)$.  The score and expected information functions, as shown in Supplementary Section~\ref{app:count-info}, are given by
\begin{eqnarray*}
 s(\bbeta;\bW) &=& \sum_{i=1}^N(W_i-\mu_i)\bx_{Ci} = \bX_C^\top(\bW-\bmu)
 \\
 I(\bbeta) &=& \sum_{i=1}^N \frac{\partial \mu_i}{\partial \bbeta^\top}\cdot \bx_{Ci} = \bX_C^\top D_C \bX_C,
\end{eqnarray*}
where $\bmu=(\mu_1,\ldots, \mu_N)$.  Note that the formula for $I(\bbeta)$ is identical to (\ref{eq:info-count}).

As for the binary case, the Fisher scoring algorithm is not guaranteed to increase the log-likelihood at a given step, unless $\bbeta$ is close enough to the maximum $\hbbeta$, or unless a step-control version $\bbeta \to \bbeta + \epsilon I(\bbeta)^{-1}\cdot s(\bbeta;\bW)$ is used, for a small enough $\epsilon$.  In practice, the EM algorithm can be used to get to a solution in vicinity of the MLE, followed by Fisher scoring steps to achieve fast convergence \citep{PinheiroBates2000,BatesPinheiro1998}.  Further details are given in section~\ref{sec:simul}.

\subsection{An overdispersed Poisson model}

The Poisson assumption is known to be often restrictive in practice, due to omitted covariates and clustering
\citep{McCullaghNelder1989,Cox1983}
The quasi-Poisson extension for $W_{ki}, k=0,1, i=1,\ldots, N$ retains the mean relationship $\mu_{ki} = G(\eta_{ki})$, 
but relaxes the variance assumption to $\Var(W_{ki}) = \phi\mu_{ki}$, for overdispersion parameter $\phi > 0$ \citep{Wedderburn1974,VerHoefBoveng2007}.  From independence of $W_{0i}, W_{1i}$ it follows that $\mu_i = \mu_{0i}+\mu_{1i}$ and $\Var(W_i) = \phi\mu_i$.  The estimation of $\bbeta$ remains unchanged, with the relevant log-likelihood, score, and information functions receiving an additional factor $\phi^{-1}$.  The MLE $\hbbeta$ can be obtained via either EM or Fisher scoring approaches, or a combination of the two. 
The overdispersion parameter $\phi$ can be estimated from the Pearson residuals as 
\[
 \hat\phi=\frac{1}{N-p}\sum_{i=1}^N
 \frac{(W_i-\hmu_i)^2}{\hmu_i},
\]
where $\hmu_i = \mu_i(\hbbeta)$, and $p$ is the dimension of $\bbeta$ \citep{McCullaghNelder1989}.  The modified estimator of $\Var(\hbbeta)$ is
\begin{equation*}
\widehat{\Var}(\hbbeta)=\hat\phi\,I(\hbbeta)^{-1},
\end{equation*}
where $I(\bbeta)$ is given by (\ref{eq:info-count}).

\section{Simulation Study and Software} \label{sec:simul}

The R package \textit{maskcave} implements the binary and count models of Sections~\ref{sec:methods} and~\ref{sec:counts}. For binary outcomes it fits the complementary log-log, log and logit links; for counts, it fits the Poisson model, with the optional quasi-Poisson adjustment of Section~\ref{sec:counts}. The MLE is computed by EM, EMG or direct Fisher scoring, with step-halved Fisher scoring as the default. The package reports standard errors, the relative exposure slope $\rho$, and, for each subject, the conditional expected numbers of events separately for treatment and control conditions, given the observed outcome.

\subsection{Design and performance measures}\label{sec:simulation_design}

We evaluated the binary outcome model in 72 settings, each with 10,000 simulated datasets without covariates (Supplementary Table~\ref{tab:sim_all72}). Total exposure was $h_i=1$. Each subject had probability 0.85 of some treatment exposure; the remaining subjects had only control exposure ($h_{1i}=0$), and $h_{0i}=1-h_{1i}$. Given the exposures, $W_{0i}$ and $W_{1i}$ were independent Poisson counts with means $h_{0i}\lambda_0$ and $h_{1i}\lambda_1$, where $\lambda_0=\exp(\alpha_0)$ and $\lambda_1=\exp(\alpha_0+\alpha_1)$, and we observed $Y_i=\min(W_{0i}+W_{1i},1)$.

We set the overall event probability $p=P(Y_i=1)$ at 1\%, 5\% or 15\%, the rate ratio $R=\exp(\alpha_1)$ equal to 0.5, 1, 2 or 7, and $N$ equal to 1,000 or 10,000. Treatment exposure among subjects with $h_{1i}>0$ followed $h_{1i} \sim \mathrm{Beta}(1,30)$ or $\mathrm{Beta}(0.1,3)$ distributions, with the same mean, 3.2\%, but standard deviations of 3.1\% and 8.7\%, or the longer $\mathrm{Beta}(2,5)$, with mean 28.6\% and standard deviation 16.0\%. For each combination of $p$, $R$ and exposure distribution, $\alpha_0$ was chosen so that the overall event probability averaged over the exposure distribution, equalled $p$. 

Under the complementary log-log link without covariates, the log-likelihood is concave in $(\lambda_0,\lambda_1)$, because each $\mu_i=\lambda_0h_{0i}+\lambda_1h_{1i}$ is linear in them. Its maximum over $\lambda_0,\lambda_1\ge0$ either has both rates positive, with finite $\halpha_1$, or lies on an edge, $\hat{\lambda}_0=0$ or $\hat{\lambda}_1=0$, so $\halpha_0=-\infty$ or $\halpha_1=-\infty$ \citep{VardiSheppKaufman1985,Marschner2010}. We call the latter situation boundary datasets.

The simulation datasets were fitted under the complementary log-log link by EMG with step halving on $Q$ and by direct Fisher scoring. (Additional simulation results using log and logit links for the fit are including in the supplementary materials.) We assessed the percentage of boundary datasets and, over valid fits, the bias, root mean squared error (RMSE) and coverage of 95\% Wald intervals for $\alpha_1$. A fit was valid if it met the stopping rule and returned finite estimates, a finite log-likelihood and a positive definite covariance matrix from the observed information. Because these measures exclude boundary datasets, in which $\halpha_1$ is infinite, they may be misleading when boundary datasets are frequent \citep{MorrisWhiteCrowther2019} and should be read together with the boundary percentage.

\subsection{Binary model results}\label{sec:simulation_results}

\begin{figure}[!tbp]
\centering
\includegraphics[width=\textwidth]{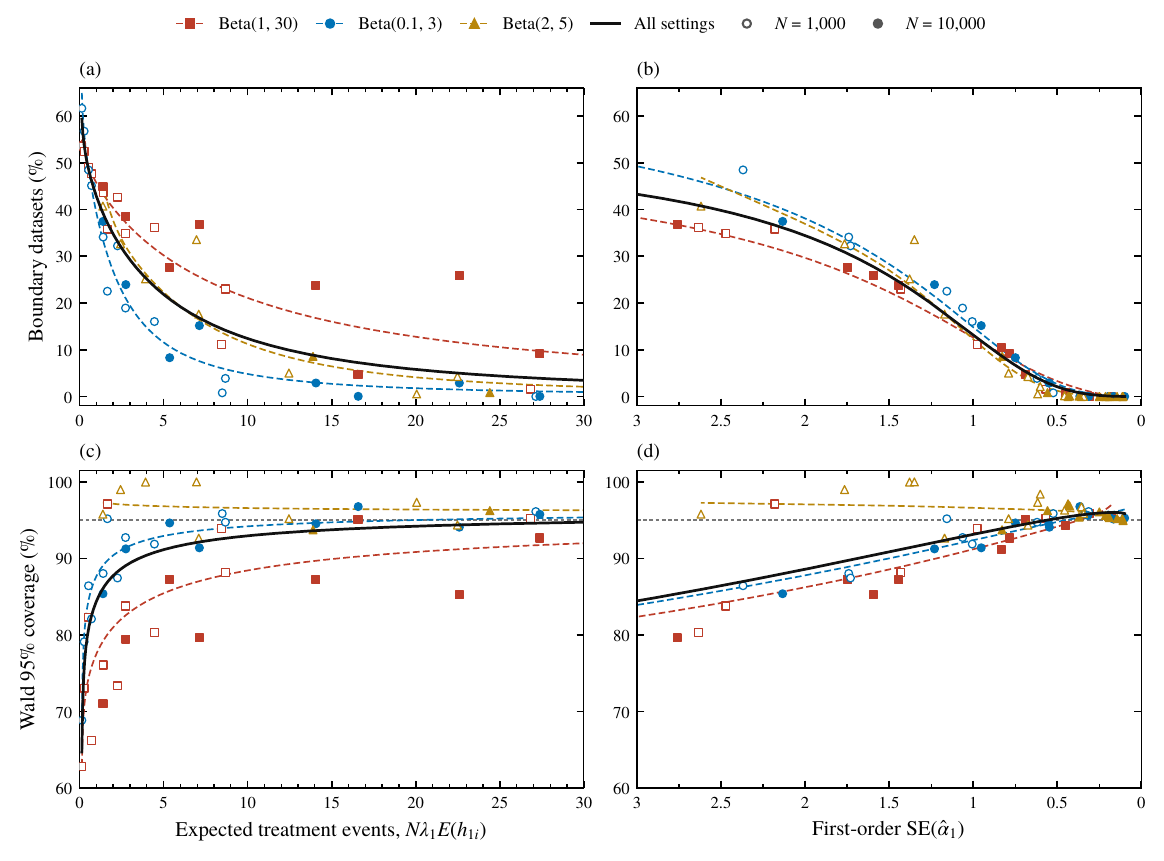}
\figwidth=\textwidth
\caption{Binary model: boundary datasets (a, b) and 95\% Wald coverage (c, d), against expected events during treatment exposure (left) and the first-order SE from expected information at the generating parameters, averaged over the exposure distribution (right; reversed axis). Each point represents a setting with 10,000 datasets. Left panels show settings with up to 30 expected treatment events; right panels show settings with first-order SE up to 3. Open symbols, $N=1{,}000$; filled symbols, $N=10{,}000$. Dashed curves are penalized spline smooths by exposure distribution; the solid black curve pools all settings. All smooths use the full set of 72 settings. Smoothing uses log x and binomial responses. The grey reference line marks 95\% coverage. Coverage uses valid fits only. Curves are descriptive and not adjusted for $p$, $N$ or $R$}
\label{fig:sim_smooth}
\par\smallskip
{\footnotesize\noindent
\textit{Alt text:} Four panels show boundary percentages and Wald coverage for the binary model against expected treatment events and first-order standard error, with the latter axis reversed. Boundary percentages generally fall as expected treatment events increase or standard errors decrease. At similar expected treatment event counts, boundary percentages are generally higher and coverage lower under Beta(1,30) exposure. Boundary curves align more closely against standard error. Coverage approaches 95\% at small standard errors; with less information, coverage can be above nominal under Beta(2,5) exposure and below nominal under the two short-exposure distributions. Coverage is computed over valid fits.
\par}
\end{figure}

Figure~\ref{fig:sim_smooth} summarizes EMG performance as a function of the expected number of events during treatment exposure, $N\lambda_1\E(h_{1i})$ (left panels), and of the first-order standard error of $\halpha_1$ (right panels). The latter is $\{[N\E_h\{I_i(\bbeta)\}]^{-1}_{\alpha_1\alpha_1}\}^{1/2}$, evaluated at the generating parameters, with the per-subject expected information averaged over the exposure distribution. 
Results for every setting, including RMSE, bias and numbers of valid fits, are given in Supplementary Table~\ref{tab:sim_all72}. 

As expected events during treatment exposure increased, coverage approached 95\%, and boundary percentages and RMSE generally decreased 
(Figure~\ref{fig:sim_smooth}a, c; Supplementary Table~\ref{tab:sim_all72}). Among the four settings with 80--100 expected treatment events, boundary percentages were at most 0.8\%, the maximum absolute bias was 0.101, and coverage ranged from 94.3\% to 95.9\% (Supplementary Table~\ref{tab:sim_all72}). Thus, asymptotic Wald intervals achieved near-nominal coverage with fewer than 100 expected treatment events in these settings. With fewer than 10 expected treatment events, boundary percentages reached 61.7\% and coverage was as low as 62.8\%.

The two short-exposure distributions had the same mean exposure and similar expected treatment event counts. At matched $N$, $p$ and $R$, coverage was consistently lower under $\mathrm{Beta}(1,30)$ than under $\mathrm{Beta}(0.1,3)$; RMSE and boundary percentages were generally higher (Figure~\ref{fig:sim_smooth}a, c). For example, with $N=10{,}000$, $p=5\%$ and $R=1$, both distributions gave 14.1 expected treatment events. Under $\mathrm{Beta}(0.1,3)$ and $\mathrm{Beta}(1,30)$, respectively, coverage was 94.5\% and 87.2\%, boundary percentages were 2.9\% and 23.8\%, and RMSE was 0.722 and 0.978 (Supplementary Table~\ref{tab:sim_all72}).

Near-nominal coverage was also attained at the smaller sample size $N=1{,}000$. Under $\mathrm{Beta}(2,5)$ exposure with $R=7$, increasing $p$ from 1\% to 15\% increased expected treatment events from 7.0 to 114.2. The boundary percentage fell from 33.6\% to 0.01\%, RMSE decreased from 1.325 to 0.363, and coverage approached the nominal level, decreasing from 100\% to 95.8\% (Supplementary Table~\ref{tab:sim_all72}). At $p=1\%$, the mean estimated standard error among valid fits was 2.72, compared with an empirical standard deviation of 1.27, consistent with overly wide Wald intervals. Thus, sufficient event and exposure information yielded near-nominal coverage even with $N=1{,}000$.

Against the first-order standard error, the pattern was the same: as the first-order standard error decreased, boundary percentages fell and coverage approached 95\% (Figure~\ref{fig:sim_smooth}(b, d)). With a first-order standard error of at most 0.5, boundary percentages were at most 0.8\% and coverage ranged from 94.3\% to 97.1\%; above 1, boundary percentages ranged from 16.0\% to 61.7\%. The boundary percentages of the three exposure distributions followed nearly the same curve against the first-order standard error, but not against expected treatment events (Figure~\ref{fig:sim_smooth}a, b), because the first-order standard error accounts for the spread of treatment exposure as well as the number of events. At larger first-order standard errors, coverage declined for the two short exposure distributions, whereas under $\mathrm{Beta}(2,5)$ it ranged from 92.6\% to 100\% (Figure~\ref{fig:sim_smooth}d).

\subsection{Count model results}\label{sec:simulation_counts}
We evaluated the count model in a separate set of 72 settings, each with 10,000 simulated datasets (Supplementary Table~\ref{tab:sim_count_all72}). Exposures and event counts were generated as in Section~\ref{sec:simulation_design}, with the same three exposure distributions and rate ratios, and the counts $W_i=W_{0i}+W_{1i}$ were analyzed with the Poisson model. We set $N$ equal to 100 or 1,000 and the mean count per subject, $\E(W_i)$, equal to 0.3, 1 or 3, with $\lambda_0$ chosen to give this mean for each exposure distribution and $R$. Among subjects with at least one event, 14--16\%, 42--44\% and 80--84\% had two or more events at these three means. Settings with $N=100$ and $\E(W_i)=3$ have the same expected numbers of events as those with $N=1{,}000$ and $\E(W_i)=0.3$. The count log-likelihood is also concave in $(\lambda_0,\lambda_1)$, and boundary datasets were defined as for the binary model; for the Poisson model they can be identified in closed form (Supplementary Section~\ref{app:count-72}). The datasets were fitted by direct Fisher scoring with step halving, valid fits were defined as in Section~\ref{sec:simulation_design}, and Wald intervals used the observed information; with the expected information, coverage differed by at most 0.8 percentage points.

\begin{figure}[!tbp]
\centering
\includegraphics[width=\textwidth]{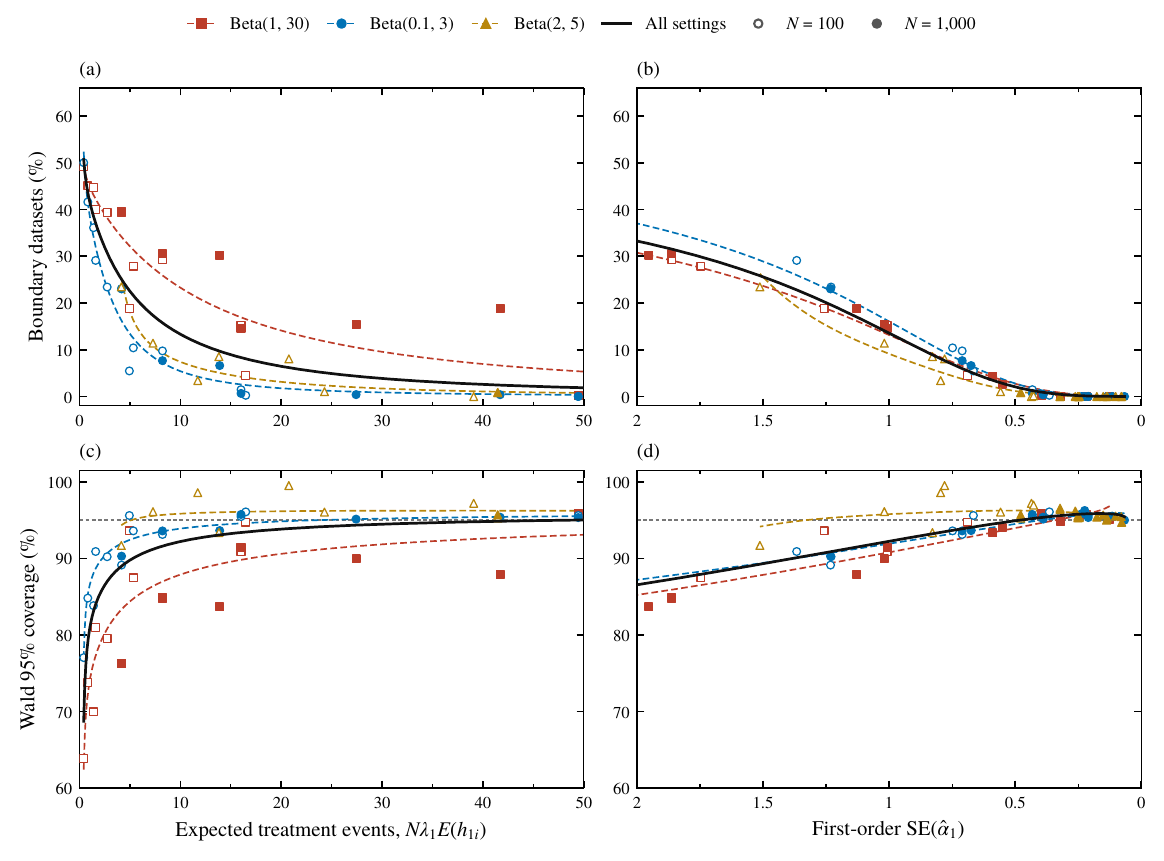}
\figwidth=\textwidth
\caption{Count model: boundary datasets (a, b) and 95\% Wald coverage (c, d), against expected events during treatment exposure (left) and the first-order SE of $\widehat\alpha_1$ from the expected information of the count model at the generating parameters, averaged over the exposure distribution (right; reversed axis), for the 72 count settings of Section~\ref{sec:simulation_counts}. Each point represents a setting with 10,000 datasets. Left panels show settings with up to 50 expected treatment events; right panels show settings with first-order SE up to 2. Open symbols, $N=100$; filled symbols, $N=1{,}000$. Dashed curves are penalized spline smooths by exposure distribution; the solid black curve pools all settings. All smooths use the full set of 72 settings. The grey reference line marks 95\% coverage. Coverage uses valid fits only. Curves are descriptive and not adjusted for $\E(W_i)$, $N$ or $R$}
\label{fig:count_smooth}
\par\smallskip
{\footnotesize\noindent
\textit{Alt text:} Four panels show boundary percentages and Wald coverage for the count model against expected treatment events and first-order standard error, with the latter axis reversed. Boundary percentages generally fall as expected treatment events increase or standard errors decrease. At similar expected treatment event counts, Beta(1,30) exposure generally gives more boundary datasets and lower coverage than the other exposure distributions. Differences between boundary curves are smaller when plotted against standard error. Coverage approaches 95\% at small standard errors, while some settings under Beta(2,5) exposure have above-nominal coverage. Coverage is computed over valid fits.
\par}
\end{figure}

Overall, 11.6\% of the count datasets were boundary datasets. Boundary percentages fell and coverage approached 95\% as expected treatment events increased or the first-order standard error decreased (Figure~\ref{fig:count_smooth}). In the 35 settings with a first-order standard error of at most 0.5, boundary percentages were at most 1.5\%, the maximum absolute bias was 0.114, and coverage ranged from 94.7\% to 97.2\%. With fewer than 10 expected treatment events, boundary percentages reached 50.0\% and coverage was as low as 63.8\%. Under $\mathrm{Beta}(2,5)$ exposure with $N=100$ and $\E(W_i)=0.3$, coverage was above nominal, up to 99.5\%. With about 30 expected events in total, valid fits near the boundary had large standard errors and hence conservative intervals; coverage fell to 95.4\% when $\E(W_i)=3$. Settings with the same expected numbers of events performed similarly: between $N=100$ with $\E(W_i)=3$ and $N=1{,}000$ with $\E(W_i)=0.3$, boundary percentages differed by at most 2.1 percentage points and coverage by at most 1.2 percentage points.

To examine overdispersion, we used the six settings with $N=1{,}000$, $R=2$ and $\E(W_i)=1$ or 3, and generated counts with $\phi=1.5$ or 3 under two mechanisms, with 10,000 datasets for each combination (Supplementary Table~\ref{tab:count_overdispersion}). In the first, the counts $W_{ki}$ were generated independently from Poisson distributions with gamma-distributed means, with shape $\mu_{ki}/(\phi-1)$ and scale $\phi-1$. This gives $\E(W_{ki})=\mu_{ki}$ and $\Var(W_{ki})=\phi\mu_{ki}$, the quasi-Poisson model of Section~\ref{sec:counts}. In the second, the two counts of a subject shared a gamma-distributed multiplier with mean 1, so that the variance-to-mean ratio of $W_i$ varied between subjects and was $\phi$ on average. With $\phi=1.5$ or 3, Wald intervals based on the Poisson information covered $\alpha_1$ in only 70.5--91.6\% of the datasets. The Pearson quasi-Poisson adjustment of Section~\ref{sec:counts} restored coverage to 93.4--95.9\%, under both mechanisms, except under $\mathrm{Beta}(1,30)$ exposure with $\E(W_i)=1$, where overdispersion raised the boundary percentage to 5.8--13.8\% and coverage was 90.9--93.1\%. With $\phi=1.5$, below the popular rule-of-thumb threshold $\hat\phi>2$, Poisson coverage was already 88.5--91.6\%. Among valid fits, a Pearson $\chi^2$ test using $N-2$ degrees of freedom and a 5\% significance level rejected $\phi=1$ in 100\% of datasets generated with $\phi=1.5$ or 3, compared with 5.0--5.3\% of those generated with $\phi=1$ (Supplementary Table~\ref{tab:count_overdispersion}). This agrees with \citet{PayneEtAl2018}, who examined thresholds for overdispersion under the Poisson model, and recommended values such as 1.5.

\subsection{Comparison of computational methods}\label{sec:simulation_computation}

Direct Fisher scoring with step halving provides an alternative to EMG for maximizing the observed-data likelihood. In the 72 simulation settings of Section~\ref{sec:simulation_design}, the two methods gave the same fits within numerical precision: among the 592,226 datasets in which both fits were valid, estimates of $\alpha_1$ differed by at most $2\times10^{-5}$ and log-likelihoods by at most $5\times10^{-10}$. EMG reached its 20,000-iteration limit in 482 of the 594,296 non-boundary datasets (0.08\%). Direct Fisher scoring gave valid fits for 471 of these datasets, after a median of 18 iterations. These estimates were finite but close to the edge $\lambda_1=0$, with median $\halpha_1=-5.2$ and median SE 380, giving very wide Wald intervals.

Given a dataset, all competing MLE algorithms maximize the same log-likelihood, so we compared their speed and their behavior with usual and poor starting values. To compare computation time under a common stopping rule \citep{Proust2017lcmm} and a common limit of 20,000 updates, we used 1,000 binary and 1,000 count datasets with finite MLEs, 250 from each of four binary settings in Table~\ref{tab:sim_all72} and each of the four count scenarios (Supplementary Section~\ref{app:timing} and Table~\ref{tab:timing_common}). Every algorithm converged in all datasets. All converged fits reached the same maximum within numerical precision, which agreed with an independently computed maximum. Direct Fisher scoring with step halving was fastest in every setting. For the binary model, median times were 0.013 seconds for direct Fisher scoring, 0.14 seconds for EM with Newton--Raphson M-steps and 0.11 seconds for EMG with one Fisher scoring step per M-step. The median time ratios to direct Fisher scoring on the same dataset were 10 and 8 for EM and EMG, respectively. For counts, median times were 0.011 seconds for direct Fisher scoring, 0.11 seconds for EM with iterated M-steps and 0.066 seconds for EM with a closed-form M-step. 

To assess the role of step halving from poor starting values, we decreased the usual starting value of $\alpha_0$ by 10, leaving $\alpha_1$ unchanged, in the first 25 datasets of each setting. With step halving on the log-likelihood for direct Fisher scoring and on $Q$ for EM and EMG, all methods, including the hybrids, reached the same maximum in the 100 binary and 100 count datasets, and direct Fisher scoring remained fastest. Without step halving, direct Fisher scoring and all tested EM and EMG variants with numerical M-steps failed at the first update. Count EM with a closed-form M-step converged in all 100 datasets without step halving.

\section{Application to the ABCD study}
\label{sec:application_primary}

\subsubsection{Baseline mTBI history and sports participation}

We applied the binary model to ABCD Release 7.0 baseline data from 10,948 children aged 9--10 years, including 9,317 (85.1\%) who participated in at least one sport, and 423 children with an mTBI history.  The baseline questionnaire records participation in each of 23 sports or activities, and weekly time spent in each of them. The outcome $Y_i$ was a parent-reported history of mTBI. A minority of children with moderate to severe traumatic brain injury or an unclassifiable injury history were excluded. Injury histories were obtained from the parent-reported traumatic brain injury screen \citep{Barch2018}; children with no injury formed the comparison group. Whether mTBI occurred during sports, $Y_{1i}=1$, or during other activities, $Y_{0i}=1$, was not recorded.

For clarity, let treatment $Z_i$ be for the moment participation in any sports (i.e., all 23 sports combined).  We assume $h_i$, the total daily wake time, to be 16 hours, common to all children. The daily exposure to sports, $h_{1i}$, was derived from the questionnaire data.  The non-sports exposure was taken as $h_{0i}=h_i-h_{1i}$. For the 14.9\% of children with no sports participation, $h_{0i}=h_i$, $h_{1i}=0$, $Y_{0i}=Y_i$, and $Y_{1i}=0$.

The daily sport exposure $h_{1i}$ was computed in two ways: the \textit{average} across reported sports exposures, and the \textit{maximum} of all sport-specific daily exposure times. Sports without valid time information were excluded from the average. These summaries allow for the possibility that children participated in different sports at different times.  Maximum daily sport exposure was 3.5 hours.

All included children had complete outcome and covariate data.  Models adjusted for age, sex, race and ethnicity, parental education, household income, and the Child Behavior Checklist internalizing and externalizing T-scores. 

We fitted each complementary log-log model by EMG and direct Fisher scoring, both with step halving; the two algorithms gave the same estimates. The log and logit models were fitted by direct Fisher scoring with step halving from four starting values, all of which gave the same estimates.

The underlying conditional rate ratio of events, for sports versus non-sports activities, was $\exp(\halpha_1) = 6.96$, 95\% CI $(2.15, 22.54)$ using average sports exposure, and $\exp(\halpha_1) = 8.78$, 95\% CI $(4.47, 17.27)$ using maximum sports exposure. The corresponding estimates of $\rho$ were 0.373 (0.072--1.346) and 0.487 (0.217--1.017) per reported hour/day: at fixed covariates, each additional daily hour of sport participation increased the expected injury count by 37\% and 49\% of the count with no sport (expected-count ratios $\Lambda(1)/\Lambda(0)= 1.373$ and $1.487$).

The maximum exposure model had better data support than the average exposure model,  Akaike information criterion (AIC) 3481.4 versus 3490.0.  
The corresponding complementary log-log, log, and logit links had nearly identical AIC values  (Table~\ref{tab:anysport_mean_max}). With average exposure, for log link the conditional risk ratio was $\exp(\halpha_1)=7.109$ $(2.260, 22.366)$, and for logit link the odds ratio was $\exp(\halpha_1)=6.812$ $(2.043, 22.714)$; with maximum exposure these values were $\exp(\halpha_1)=8.837$ $(4.580, 17.054)$ for log link and $8.722$ $(4.353, 17.476)$ for logit link. For each exposure calculation, the AIC values of the three links differed by less than 0.6, and the expected numbers of children injured in each setting by less than 0.3. The fitted event probabilities were below 0.25, so the three links nearly coincide over the observed exposures, as in the simulations (Supplementary Table~\ref{tab:sim_links72}), and the corresponding rate ratio, risk ratio, and odds ratios are close approximations of each other.

The formulas in Section~\ref{sec:em} provide the expected number of children with an mTBI history in each injury setting given the data: sport only, non-sport only, or both (Table~\ref{tab:anysport_mean_max}). Of the 423 children with an mTBI history, 373 were sport participants, but only an estimated 57.6 and 96.9 children, under average and maximal exposures respectively, had at least one mTBI during sport.

\begin{table}[!tbp]
\begin{singlespace}
\centering\footnotesize
\setlength{\tabcolsep}{5pt}
\caption{Baseline mTBI analysis: adjusted estimates under two summaries of daily sport exposure.}
\label{tab:anysport_mean_max}
\begin{tabular}{@{}lcccc@{}}
\toprule
 & \multicolumn{2}{c}{\makecell{Exposure measured by the\\average across reported sports}} & \multicolumn{2}{c}{\makecell{Exposure measured by the\\largest sport-specific value}} \\
\cmidrule(lr){2-3}\cmidrule(l){4-5}
 & Estimate (95\% CI) & AIC & Estimate (95\% CI) & AIC \\
\midrule
\multicolumn{5}{@{}l}{\textit{Exposure association, $\exp(\alpha_1)$, by link}} \\
\quad Cloglog: rate ratio & 6.960 (2.149--22.540) & 3490.0 & 8.784 (4.467--17.274) & 3481.4 \\
\quad Log: risk ratio & 7.109 (2.260--22.366) & 3490.3 & 8.837 (4.580--17.054) & 3481.5 \\
\quad Logit: odds ratio & 6.812 (2.043--22.714) & 3489.7 & 8.722 (4.353--17.476) & 3481.2 \\
\addlinespace[4pt]
\multicolumn{5}{@{}l}{\textit{Relative exposure slope, cloglog}} \\
\quad \makecell[l]{$\rho$: change in expected injury count per\\additional sport hour/day, relative to no sport} & 0.373 (0.072--1.346) & & 0.487 (0.217--1.017) & \\
\addlinespace[4pt]
\multicolumn{5}{@{}l}{\textit{Expected mTBI case by injury setting, cloglog}} \\
\quad Sport only & 55.2 (20.0--131.8) & & 93.3 (57.1--143.4) & \\
\quad Non-sport only & 365.4 (286.6--402.0) & & 326.1 (275.0--363.4) & \\
\quad Both settings & 2.5 (1.1--5.5) & & 3.6 (2.6--5.0) & \\
\quad Observed total & 423 & & 423 & \\
\bottomrule
\end{tabular}
\par\vspace{4pt}{\footnotesize\raggedright\setstretch{1.15}
All models use the same 10,948 children and adjust for age, sex, race and ethnicity, parental education, household income, and CBCL internalizing and externalizing T-scores. Female and White were the reference categories for sex and race and ethnicity, respectively. Age, parental education and household income categories, and CBCL T-scores entered linearly; education and income used the most recent values available up to baseline. Total waking time is fixed at 16 hours/day. Under the log and logit links, $\exp(\alpha_1)$ is the sport-to-non-sport risk ratio and odds ratio, respectively, at equal exposure. Estimates of $\rho$ and expected case counts use the cloglog link. Expected case counts sum the conditional probabilities from the E-step over the 423 children with an mTBI history (Section~\ref{sec:em}); the three categories are mutually exclusive and sum to 423 before rounding. Effect CIs use the observed information. The confidence limits for $\rho$ are obtained by subtracting 1 from each confidence limit for the rate ratio $\exp(\alpha_1)$ and dividing by 16. Case-count CIs use the delta method on the logit scale for proportions of these 423 children.

Abbreviations: cloglog, complementary log--log; CI, confidence interval; AIC, Akaike information criterion; CBCL, Child Behavior Checklist; mTBI, mild traumatic brain injury.
\par}
\end{singlespace}
\end{table}

\subsubsection{Assumed waking time}

Under the complementary log-log model, a different assumed waking time changes only the rate-ratio scale of these results. By Theorem 1, $\hat\rho$, the fitted probabilities and AIC are unchanged, and $\exp(\halpha_1)=1+h\hat\rho$. Under average exposure assumptions, for example, the conditional rate ratio is 3.98, 6.96 and 9.94 at 8, 16 and 24 waking hours/day; under maximum exposure it is 4.89, 8.78 and 12.68. Because daily sport exposure was at most 3.5 hours and both slopes are positive, the conditions of Section~\ref{sec:extensions} hold at all three waking times.

\subsubsection{Year 5 incident concussions and sports participation}

Starting with the Year 5 visit, parents in the ABCD study were asked to report the number of concussions their child had during sport or physical activity, since the previous yearly assessment. Counts were coded from 0 to 4 or more. We required a known count, a matching sports questionnaire, at least one reported activity, complete activity time information and a positive reporting interval.

We used the count model of Section~\ref{sec:counts} to evaluate the concussion risk of individual sports, relative to the other sports that children played, among children engaged in sports. For each of the ten most commonly reported sports, a separate model took that sport as treatment condition,  compared to all other reported sports combined, as control. Exposures $h_{ki}$ were reported as daily hours multiplied by the number of days since the previous assessment, so total exposure varied across children. The analysis included 6,379 children with complete covariates, with 304 concussions recorded in 272 children. Covariates were measured at Year 5, with Year 4 CBCL scores used when Year 5 scores were missing.

The quasi-Poisson scale parameter was 1.76--1.89 in the fitted models, suggestive but not compelling evidence of overdispersion \citep{PayneEtAl2018}. Table~\ref{tab:application_count} gives both Poisson and quasi-Poisson intervals (Section~\ref{sec:counts}). The adjusted rate ratio, comparing a sport with the other sports per hour of play, was 2.46 for football (quasi-Poisson 95\% CI, 1.37--4.43) and 2.41 (1.45--4.00) for soccer. For track, running and cross-country, the rate ratio of 0.15 had a Poisson confidence interval that excluded the value 1, (0.024--0.939), but a quasi-Poisson interval that did not (0.013--1.709); the quasi-Poisson intervals for the other seven sports also included 1. The binary model, fitted to the indicators $Y_i=\ind(W_i>0)$ as observed outcomes on the same dataset, gave similar rate ratios, for example 2.69 (1.79--4.04) for football and 2.40 (1.65--3.50) for soccer. EMG and direct Fisher scoring algorithms reached identical MLE in all 20 count models.

\begin{table}[!tbp]
\begin{singlespace}\centering\footnotesize\setlength{\tabcolsep}{4pt}
\caption{Year 5 concussion rates: adjusted comparisons of each sport with the other reported sports.}\label{tab:application_count}
\begin{tabular}{@{}lrrccccc@{}}
\toprule
 & & & \multicolumn{3}{c}{Count model} & \multicolumn{2}{c}{Binary model} \\
\cmidrule(lr){4-6}\cmidrule(l){7-8}
 & & & & \multicolumn{2}{c}{95\% CI} & & \\
\cmidrule(lr){5-6}
Sport & \makecell{Players,\\$n$ (\%)} & \makecell{Average sport\\exposure\\(hours/day)} & \makecell{Rate\\ratio} & Poisson & \makecell{Quasi-\\Poisson} & \makecell{Rate\\ratio} & 95\% CI \\
\midrule
Basketball & 1,618 (25.4) & 0.82 & 0.876 & 0.477--1.611 & 0.383--2.003 & 0.799 & 0.438--1.457 \\
Soccer & 1,234 (19.3) & 0.81 & 2.409 & 1.656--3.504 & 1.452--3.998 & 2.399 & 1.647--3.495 \\
\makecell[l]{Track / running /\\cross-country} & 1,160 (18.2) & 0.95 & 0.151 & 0.024--0.939 & 0.013--1.709 & 0.160 & 0.033--0.770 \\
Football & 1,005 (15.8) & 1.11 & 2.463 & 1.606--3.776 & 1.368--4.433 & 2.689 & 1.791--4.038 \\
Volleyball & 945 (14.8) & 0.79 & 0.928 & 0.479--1.795 & 0.378--2.274 & 1.021 & 0.569--1.833 \\
\makecell[l]{Swimming /\\water polo} & 900 (14.1) & 0.79 & 0.702 & 0.321--1.536 & 0.243--2.025 & 0.695 & 0.303--1.594 \\
Baseball / softball & 711 (11.1) & 0.98 & 0.428 & 0.139--1.319 & 0.093--1.969 & 0.474 & 0.187--1.203 \\
Skateboarding & 655 (10.3) & 0.53 & 0.379 & 0.074--1.947 & 0.041--3.485 & 0.444 & 0.126--1.571 \\
\makecell[l]{Skiing,\\snowboarding} & 637 (10.0) & 0.78 & 0.872 & 0.319--2.384 & 0.219--3.465 & 0.757 & 0.260--2.200 \\
Ballet / dance & 560 (8.8) & 0.87 & 0.487 & 0.188--1.265 & 0.134--1.776 & 0.533 & 0.217--1.310 \\
\midrule
\makecell[l]{Total, any of\\the ten sports} & 5,364 (84.1) & 1.50 & & & & & \\
\bottomrule
\end{tabular}
\par\vspace{4pt}{\footnotesize\raggedright\setstretch{1.15}Each row compares the named sport with all other reported sports in a separate model, including children with no exposure to that sport and those reporting only that sport. The rate ratio compares concussion rates per hour of play. All models use 6,379 children with complete covariates, of whom 272 reported 304 concussions. Players are given as a percentage of these children. Average sport exposure is the participants\textquotesingle{} mean daily time in that sport. The total row counts each child once; its exposure is the child\textquotesingle{}s daily time summed over the ten sports. Exposure is daily hours multiplied by days since the previous assessment, with covariates as in Table~\ref{tab:anysport_mean_max}. Count-model rate-ratio intervals use expected information, with covariance multiplied by the Pearson dispersion $\widehat\phi$ (1.76--1.89) for quasi-Poisson intervals. The binary cloglog model uses $Y_i=\ind(W_i>0)$, direct Fisher scoring with step halving and observed-information intervals. Intervals are not adjusted for multiple comparisons.
\par}
\end{singlespace}
\end{table}

\section{Discussion}

We proposed a statistical model for outcomes representing events or counts of events occurring under two or several treatment conditions, when exposure time differs between conditions and between individuals, and the cause of each event is masked, i.e., not directly attributed to one of the conditions. The variation in exposure times allows the quantification of relative risk between treatment conditions.  Maximum likelihood estimation is implemented using either an EM algorithm, with the causes of events as missing data, or by direct maximization using Newton-Raphson or Fisher scoring algorithms.  This development was motivated by a study of the risk of brain injury associated with sports activities in the ABCD study.  The binary outcome model is justified via underlying independent Poisson count processes over each treatment condition, which lead to separate GLM's for each treatment condition, with additive complementary log probabilities.  While the complementary log-log link for each GLM has a natural justification, the link function specification is general. In particular, complementary log-log, log, and logit links allow simple, attractive interpretations for the treatment effect, as conditional, or adjusted, rate ratio, risk ratio, and odds ratio, respectively.

We implemented the models and algorithms, under three usual options for the link function, in an R package, \textit{maskcave}.  We studied the performance of the model and computational algorithms via extensive simulations.  The MLE asymptotics apply when the expected numbers of events under each treatment conditions are moderate or large.  The performance deteriorates, as expected, when these numbers become small, or when the variation in exposure across individuals is small.
When exposure varies little between subjects, the data mainly inform the overall event rate, and its division between treatment conditions is less well determined. As simulations show, a larger sample size alone did not ensure good performance when treatment events were rare. In planning a study, the first-order standard error can be computed from the planned sample size, exposure distribution and assumed event rates, and is a useful guide together with the expected number of treatment events.  These features apply to both binary and count outcome models.  For count data, overdispersion made Poisson intervals too narrow, and Pearson scaling restored near-nominal coverage except in settings with frequent boundary datasets. Already at $\phi=1.5$, below the common rule-of-thumb threshold $\hat\phi>2$, Poisson coverage fell below the nominal level, whereas the Pearson $\chi^2$ test detected the overdispersion in every dataset and kept close to its nominal 5\% level when $\phi=1$ (Section~\ref{sec:simulation_counts} and Supplementary Table~\ref{tab:count_overdispersion}). These results support testing for overdispersion with the Pearson $\chi^2$ test rather than relying on the cutoff $\hat\phi>2$, consistent with \cite{PayneEtAl2018}. Our simulations were limited to a gamma--Poisson data generating mechanisms; care is needed when extending conclusions to other forms of overdispersion or dependence. 

In terms of computational methods, EM, EMG and direct Fisher scoring led to identical estimates in the simulation study and in the application. In our implementation, Fisher scoring was about 10 times faster than EM and 8 times faster than EMG (EM with single Fisher scoring step at M-steps). From poor starting values, all methods except count EM with a closed-form M-step needed step halving, and with step halving all reached the same maximum.  The EM paradigm provides the expected numbers of events during each treatment condition, given the observed data.

Our application of the \textit{maskcave} model to baseline ABCD data showed that greater reported sports exposure was associated with higher risk of mTBI, after adjustment for potential demographic, child-, and family-level confounders.
In the analysis of counts of concussions during sports activities from Year 5 of ABCD study, we found that 
football and soccer had higher concussion rates per hour of play than the other sports that children played. These effects were statistically significant and clinically relevant, corresponding to about 2.5 times higher risk.  
These analyses had limitations, many of them representative of any use of this model: Adjustment for confounders in observational studies is subject to strong, unverifiable assumptions \citep{HernanRobins2020}.  In particular, we assumed a homogeneous treatment effect, that is, that covariates had the same effect under both sport and non-sport conditions.  The sports exposure and the concussions are parent-reported and subject to recall error and bias.
For the baseline ABCD analysis, we assumed a common wake time of 16 hours/day for all children;  using a different value leads to numerically different risks, while leaving statistical significance essentially unchanged.
Finally, the model assumes that the event indicators for each treatment condition are independent, given exposures and covariates.
The ABCD study had thousands of participants and hundreds of events, ensuring reliable inference.  Other applications are bound by the envelope of the method, as studied in the simulation section, and consistent with the behavior of standard Bernoulli and Poisson GLMs.

Several extensions of the \textit{maskcave} model are currently under way.  One way to account for heterogeneity between subjects is to include subject-specific random effects.  The EM algorithm setup is particularly attractive here, since treating the causes of events as known leads to a generalized linear mixed-effects model \citep{FitzmauriceLairdWare2011}.
A second extension involves applications to longitudinal data, such as those generated by the yearly visits in the ABCD study.  Within-subject correlation can be accounted for via random effects, as discussed above, or via a GEE-type M-estimator, with robust variance estimation \citep{LiangZeger1986,ZegerLiangAlbert1988}. 

The \textit{maskcave} model uses the missing data paradigm to solve a novel problem, 
with specific applications to the ABCD study, but which it
conceivably occurs in many other settings, including large observational studies and surveys. Using variation in exposure across individuals is the key piece that allows determination of treatment effects, and a novel addition to the statistical armamentarium.

\section*{Declaration of the use of generative AI and AI-assisted technologies} During the preparation of this work the authors used Claude (Anthropic) in order to check derivations and check simulation code. After using this tool the authors reviewed and edited the content as necessary and take full responsibility for the content of the publication.

\section*{Acknowledgement}
This work was supported by the Eunice Kennedy Shriver National Institute of Child Health and Human Development grant R01HD105338 (Principal Investigator: Jeffrey E. Max). We thank the ABCD participants and their families.

Data used in the preparation of this article were obtained from the Adolescent Brain Cognitive Development\texttrademark{} (ABCD) Study, held in the NIH Brain Development Cohorts Data Sharing Platform. This is a multisite, longitudinal study designed to recruit more than 10,000 children aged 9--10 and follow them over 10 years into early adulthood. The ABCD Study\textregistered{} is supported by the National Institutes of Health and additional federal partners under award numbers U01DA041048, U01DA050989, U01DA051016, U01DA041022, U01DA051018, U01DA051037, U01DA050987, U01DA041174, U01DA041106, U01DA041117, U01DA041028, U01DA041134, U01DA050988, U01DA051039, U01DA041156, U01DA041025, U01DA041120, U01DA051038, U01DA041148, U01DA041093, U01DA041089, U24DA041123, U24DA041147. A full list of supporters is available at \url{https://abcdstudy.org/about/federal-partners/}. ABCD Consortium investigators designed and implemented the study and/or provided data but did not necessarily participate in the analysis or writing of this report. This manuscript reflects the views of the authors and may not reflect the opinions or views of the NIH or ABCD Consortium investigators. The ABCD data repository grows and changes over time. The ABCD data used in this report came from NBDC Digital Object Identifier (DOI) 10.82525/8f3w-5260 \citep{ABCDReleaseDOI}.

\section*{Supplementary material}
The Supplementary Material includes derivations of the score, information, expectation--maximization updates and Louis' formula for the binary model with $K$ treatment conditions and three links, the effect of changing the total exposure, the information for the count model, and additional simulation results for the binary and count models.

\bibliographystyle{biometrika}
\bibliography{reference}

\end{document}


\raggedbottom
\renewcommand{\theequation}{\thesection.\arabic{equation}}

\maketitle

\section{Proofs and methods}\label{app:proofs-methods-n}
The derivations below are for the $K$ treatment conditions setup (section \ref{sec:extensions}), with the two-conditions setup $K=2$ as a particular case.  For individuals $i=1,\ldots, N$ we measure independent variables $Y_i$, with exposures $h_{ki}$ corresponding to treatment conditions $k=0,1,\ldots,K-1$, with $k=0$ as the control condition, and covariates $\bx_i$. The exposures must not overlap.
$Y_{ki}$ are independent binary latent (missing) outcomes during exposure to condition $k$, with $\pi_{ki}=P(Y_{ki}=1)$, and
\[
g(\pi_{ki})=\eta_{ki}=\log h_{ki}+\bbeta^\top\bx_{ki},\qquad
\bbeta=(\alpha_0,\alpha_1,\ldots,\alpha_{K-1},\bgamma^\top)^\top,
\]
where $\bx_{0i}=(1,0,\ldots,0,\bx_i^\top)^\top$ and, for $k\ge1$, $\bx_{ki}$ has ones in the positions of $\alpha_0$ and $\alpha_k$. We have $Y_i=\max_kY_{ki}$ and $1-\pi_i=\prod_{k=0}^{K-1}(1-\pi_{ki})$. Without loss of generality, we will assume $h_{ki}>0$, to avoid unnecessary notation.  Zero-exposure terms $h_{ki}=0$ are in reality ignored.
In the notation of section \ref{sec:em}, let $G=g^{-1}$ be the inverse link,
$v(\pi)=\pi(1-\pi)$ the Bernoulli variance function, and
$w_{ki}=G'(\eta_{ki})/v(\pi_{ki})$. Define $w'_{ki}=\partial w_{ki}/\partial\eta_{ki}$, $v_{ki}=v(\pi_{ki})$ and $v_i=v(\pi_i)$. 

\subsection{Direct calculation of maximum likelihood}\label{app:direct-likelihood}\label{app:information}
Let
\[
e_{ki}=\frac{G'(\eta_{ki})}{1-\pi_{ki}}=\pi_{ki}w_{ki},\qquad
f_{ki}=\frac{\partial e_{ki}}{\partial\eta_{ki}}=v_{ki}w_{ki}^2+\pi_{ki}w'_{ki},
\]
\[
\mathbf d_i=\sum_{k=0}^{K-1}e_{ki}\bx_{ki},\qquad
M_i=\sum_{k=0}^{K-1}f_{ki}\bx_{ki}\bx_{ki}^\top.
\]
Differentiating $1-\pi_i=\prod_k(1-\pi_{ki})$ gives
\[
\frac{\partial\pi_i}{\partial\bbeta}=(1-\pi_i)\mathbf d_i,\qquad
\frac{\partial\mathbf d_i}{\partial\bbeta^\top}=M_i.
\]
The score function corresponding to log-likelihood (\ref{eq:loglik}) is
\begin{eqnarray}
s(\bbeta)&=&\sum_{i=1}^N\frac{Y_i-\pi_i}{v_i}\frac{\partial\pi_i}{\partial\bbeta}
=\sum_{i=1}^N\frac{Y_i-\pi_i}{\pi_i}\mathbf d_i.\label{eq:score-K}
\end{eqnarray}
The observed information and, since $\E(Y_i)=\pi_i$, the expected information are
\begin{eqnarray}
J(\bbeta;\bY)&=&\sum_{i=1}^N\left\{\frac{Y_i(1-\pi_i)}{\pi_i^2}\mathbf d_i\mathbf d_i^\top
-\frac{Y_i-\pi_i}{\pi_i}M_i\right\},\label{eq:J-K}\\
I(\bbeta)&=&\sum_{i=1}^N\frac{1-\pi_i}{\pi_i}\mathbf d_i\mathbf d_i^\top.\label{eq:I-K}
\end{eqnarray}
Put $u_{ki}=e_{ki}/\pi_i$ and $\bx_{Bi}=\sum_k u_{ki}\bx_{ki}=\mathbf d_i/\pi_i$.
Let $\bX_B$ and $\bX_k$ have rows $\bx_{Bi}^\top$ and $\bx_{ki}^\top$, respectively, and $D_B=\diag\{v_i\}$. Then
\begin{eqnarray}
s(\bbeta)&=&\bX_B^\top(\bY-\bpi),\qquad I(\bbeta)=\bX_B^\top D_B\bX_B,\nonumber\\
J(\bbeta;\bY)&=&\bX_B^\top\diag\{Y_i(1-\pi_i)\}\bX_B
-\sum_k\bX_k^\top\diag\{(Y_i-\pi_i)f_{ki}/\pi_i\}\bX_k,\label{eq:obs-info-matrix}\\
R_J(\bbeta;\bY)&=&\bX_B^\top\diag\{(Y_i-\pi_i)(1-\pi_i)\}\bX_B
-\sum_k\bX_k^\top\diag\{(Y_i-\pi_i)f_{ki}/\pi_i\}\bX_k,\nonumber
\end{eqnarray}
where $J=I+R_J$ and $\E(R_J)=0$.
A Fisher scoring algorithm updates $\bbeta\to\bbeta+\epsilon I(\bbeta)^{-1}s(\bbeta)$.

Table~\ref{tab:link-ef} gives $e_{ki}$ and $f_{ki}$ for three commonly-used links. Under the complementary log-log link, $e_{ki}=f_{ki}=\mu_{ki}=\exp(\eta_{ki})=-\log(1-\pi_{ki})$, the mean of the latent Poisson count, and $\pi_i=1-\exp(-\mu_i)$ with $\mu_i=\sum_k\mu_{ki}$. Under the log link, $e_{ki}=\pi_{ki}/(1-\pi_{ki})$ and $f_{ki}=\pi_{ki}/(1-\pi_{ki})^2$. Under the logit link, $e_{ki}=\pi_{ki}$ and $f_{ki}=\pi_{ki}(1-\pi_{ki})$. The complete-data observed and expected information then coincide, but the observed-data ones do not: since $\pi_i=1-\prod_k(1-\pi_{ki})$, $\operatorname{logit}(\pi_i)$ is not linear in $\bbeta$, and $R_J$ is not zero unless each subject is exposed to a single condition. With $K=2$, (\ref{eq:score-K}) and (\ref{eq:I-K}) reduce to (\ref{eq:score_direct}) and (\ref{eq:info_direct}) of the main paper.

\begin{table}[!htbp]\centering
\caption{Coefficients for the observed-data score and information.}\label{tab:link-ef}
\begin{tabular}{lcc}
\toprule
Link & $e_{ki}$ & $f_{ki}$\\
\midrule
Complementary log-log & $\mu_{ki}$ & $\mu_{ki}$\\
Log & $\pi_{ki}/(1-\pi_{ki})$ & $\pi_{ki}/(1-\pi_{ki})^2$\\
Logit & $\pi_{ki}$ & $\pi_{ki}(1-\pi_{ki})$\\
\bottomrule
\end{tabular}
\end{table}

\subsection{EM algorithm}\label{app:binary-methods}
Given latent variables $Y_{ki}$, the complete-data score can be written as
$$
s_c(\bbeta;\bY_0,\ldots,\bY_{K-1}) = \sum_{i=1}^N s_i,$$
where
\begin{equation} \label{eq:si}
s_i = \sum_{k=0}^{K-1}(Y_{ki}-\pi_{ki})w_{ki}\bx_{ki}.
\end{equation}

The complete-data observed information is
\begin{eqnarray}
J_c(\bbeta;\bY_0,\ldots,\bY_{K-1})&=&-\frac{\partial s_c}{\partial\bbeta^\top}=\sum_i J_i,\nonumber\\
J_i&=&\sum_k\{v_{ki}w_{ki}^2-(Y_{ki}-\pi_{ki})w'_{ki}\}\bx_{ki}\bx_{ki}^\top,\label{eq:Ji}\\
I_c(\bbeta)&=&\sum_{k,i}v_{ki}w_{ki}^2\bx_{ki}\bx_{ki}^\top.\nonumber
\end{eqnarray}
The complete-data expected information does not include the terms with factors $(Y_{ki}-\pi_{ki})$, since their expectation is 0.

The EM algorithm alternates E- and M-steps.  At iteration $t$, given current  value $\bbeta^{(t)}=\tbbeta$,
E-step amounts to computing conditional expectations $\tY_{ki}=\E(Y_{ki}|Y_i,\tbbeta)=Y_i\tpi_{ki}/\tpi_i$, where $\tpi_i = \pi_i(\tbbeta)$, $\tpi_{ki}=\pi_{ki}(\tbbeta)$, $k=0,\ldots,K-1$; this holds because $Y_{ki}=1$ implies $Y_i=1$.

The M-step maximizes
\[
Q(\bbeta|\tbbeta)=\sum_{k,i}\{\tY_{ki}\log\pi_{ki}+(1-\tY_{ki})\log(1-\pi_{ki})\}.
\]
This is a Bernoulli quasi-log-likelihood for fractional responses $\tY_{ki}$. With these held fixed, quasi-Fisher scoring and Newton--Raphson use, respectively,
\begin{eqnarray*}
\bbeta&\to&\bbeta+\epsilon I_c(\bbeta)^{-1}s_c(\bbeta;\tilde{\bY}_0,\ldots,\tilde{\bY}_{K-1}),\\
\bbeta&\to&\bbeta+\epsilon J_c(\bbeta;\tilde{\bY}_0,\ldots,\tilde{\bY}_{K-1})^{-1}s_c(\bbeta;\tilde{\bY}_0,\ldots,\tilde{\bY}_{K-1}).
\end{eqnarray*}
Step halving on $Q$ is used when necessary. Iteration to convergence gives the full M-step; one Fisher scoring update gives EMG.

\label{app:mstep-general-links}For the complementary log-log, log and logit links, the derivative of the inverse link and the weights are given in Table~\ref{tab:link-weights}.
Substituting these weights into $s_c$, $J_c$ and $I_c$, with $\tY_{ki}$ in place of $Y_{ki}$, gives the M-step for each link.
\begin{table}[!htbp]\centering
\caption{Derivative of the inverse link and weights of the M-step for three links.}\label{tab:link-weights}
$\begin{array}{l|ccc}
 \text{Link function}& G'(\eta_{ki}) & w_{ki} & w'_{ki}\\\hline
\text{Complementary log-log} & \mu_{ki}(1-\pi_{ki}) & \mu_{ki}/\pi_{ki} & \mu_{ki}/\pi_{ki}-\mu_{ki}^2(1-\pi_{ki})/\pi_{ki}^2\\
\text{Log} & \pi_{ki} & 1/(1-\pi_{ki}) & \pi_{ki}/(1-\pi_{ki})^2\\
\text{Logit} & \pi_{ki}(1-\pi_{ki}) & 1 & 0
\end{array}$
\end{table}

Each log-link update must keep $\eta_{ki}<0$, that is $\pi_{ki}<1$, for all positive exposures. 
Since logit is the canonical link of the Bernoulli GLM, the observed and expected complete-data information coincide, so Newton--Raphson and quasi-Fisher scoring M-steps are identical.

\subsection{EM algorithm, variance calculations}\label{app:louis}
Upon convergence of EM algorithm, asymptotic variance estimates for $\var(\hbbeta)$ are
$\widehat{\var}(\hbbeta) = I(\hbbeta)^{-1}$ and $\widehat{\var}(\hbbeta) = J(\hbbeta)^{-1}$, where
$J(\bbeta)$, $I(\bbeta)$ are the observed and expected information based on $\bY$.
Expected information (\ref{eq:info}) can be derived as follows:
\begin{eqnarray*}
I(\bbeta) &=&
\var\{\E(s_c(\bbeta;\bY_0,\ldots,\bY_{K-1})\mid\bY)\}
\\&=&
\var\left(\sum_{i=1}^N Y_i\bx_{Bi}\right)
\\&=&
\sum_{i=1}^N v(\pi_i)\bx_{Bi}\bx_{Bi}^\top
\\&=& \bX_B^\top D_B \bX_B.
\end{eqnarray*}

Observed information is computed using Louis' formula (\ref{eq:obs_info}), or in the more general form, i.e., not at $\bbeta=\hbbeta$,
\begin{equation} \label{eq:louis}
J(\bbeta;\bY) = \E[J_c(\bbeta;\bY_0,\ldots,\bY_{K-1})|\bY, \bbeta] - \var(s_c(\bbeta;\bY_0,\ldots,\bY_{K-1})|\bY,\bbeta).
\end{equation}
Since $J_c, s_c$ are sum of independent components, we have that
$$
J(\bbeta;\bY) = \sum_{i=1}^N (\tJ_i - \tilde{R}_i),
$$
where $\tJ_i = \E(J_i|Y_i, \tbbeta)$, $\tilde{R}_i = \var(s_i|Y_i,\tbbeta)$, computed at $\tbbeta=\bbeta$.  Given that $s_i, J_i$ in (\ref{eq:si}, \ref{eq:Ji}) are
linear in $(Y_{0i},\ldots,Y_{K-1,i})$, the computation above only requires the first two conditional moments of $(Y_{0i},\ldots,Y_{K-1,i})$, namely
\[
\E(Y_{ki}\mid Y_i,\tbbeta)=\tY_{ki}=\frac{Y_i\tpi_{ki}}{\tpi_i},\qquad
\var(Y_{ki}\mid Y_i,\tbbeta)=\frac{Y_i\tpi_{ki}}{\tpi_i}\left(1-\frac{\tpi_{ki}}{\tpi_i}\right),
\]
\[
\Cov(Y_{ki},Y_{li}\mid Y_i,\tbbeta)=-\frac{Y_i\tpi_{ki}\tpi_{li}(1-\tpi_i)}{\tpi_i^2}\quad(k\ne l).
\]
It follows that, at $\tbbeta=\bbeta$,
\[
J(\bbeta;\bY)=\sum_{i=1}^N\Big\{\sum_k j_{ki}\bx_{ki}\bx_{ki}^\top+\sum_{k\ne l}j_{kl,i}\bx_{ki}\bx_{li}^\top\Big\},
\]
with
\[
j_{ki}=v_{ki}w_{ki}^2-\Big(\frac{Y_i\pi_{ki}}{\pi_i}-\pi_{ki}\Big)w'_{ki}
-\frac{Y_i\pi_{ki}}{\pi_i}\Big(1-\frac{\pi_{ki}}{\pi_i}\Big)w_{ki}^2,
\]
\[
j_{kl,i}=\frac{Y_i\pi_{ki}\pi_{li}(1-\pi_i)}{\pi_i^2}w_{ki}w_{li}.
\]
With $e_{ki}=\pi_{ki}w_{ki}$ and $f_{ki}=v_{ki}w_{ki}^2+\pi_{ki}w'_{ki}$, these are the coefficients of (\ref{eq:J-K}): $j_{ki}=Y_i(1-\pi_i)u_{ki}^2-(Y_i-\pi_i)f_{ki}/\pi_i$ and $j_{kl,i}=Y_i(1-\pi_i)u_{ki}u_{li}$, so the two derivations agree.

\paragraph{$K=2$ case.}
\begin{equation}
J(\bbeta;\bY)=\sum_{i=1}^N\left\{
 j_{0i}\bx_{0i}\bx_{0i}^\top
 +j_{1i}\bx_{1i}\bx_{1i}^\top
 +j_{01,i}(\bx_{0i}\bx_{1i}^\top+\bx_{1i}\bx_{0i}^\top)\right\},
 \label{eq:louis-general-closed}
\end{equation}
with
\begin{eqnarray*}
j_{0i} &=& v_i u_{0i}^2 
- (Y_i-\pi_i)\left(\frac{(1-\pi_{0i})\pi_{1i}}{\pi_{0i}}\cdot u_{0i}^2+u_{0i}
\cdot \frac{w'_{0i}}{w_{0i}}\right)
\\
j_{1i} &=& v_i u_{1i}^2 
- (Y_i-\pi_i)\left(\frac{(1-\pi_{1i})\pi_{0i}}{\pi_{1i}}\cdot u_{1i}^2+u_{1i}
\cdot \frac{w'_{1i}}{w_{1i}}\right)
\\
j_{01,i} &=& v_i u_{0i}u_{1i} + (Y_i-\pi_i)(1-\pi_i)u_{0i}u_{1i}
\end{eqnarray*}
and $j_{01,i}=j_{10,i}$ in the notation above.

\FloatBarrier
\subsection{Derivations for the relative exposure slope $\rho$}\label{app:waking}

\paragraph{Changing the assumed total exposure: complementary log-log link}\label{app:waking-cloglog}
By Theorem~\ref{thm:common-exposure} of the main paper, the event probabilities $\pi_i$, and hence the likelihood, depend on $\alpha_0$, $\alpha_1$ and $h$ only through $\log h+\alpha_0$ and $\rho$. For another total $h'$, the change of parameters
\begin{equation}
 \alpha_0'=\alpha_0+\log(h/h'),\qquad \alpha_1'=\log(1+h'\rho),\qquad \bgamma'=\bgamma,
 \label{eq:h-map}
\end{equation}
keeps $\log h+\alpha_0$ and $\rho$ fixed, and therefore every $\pi_i$ and the likelihood. It requires $h_{1i}\le h'$ for every subject, so that the control exposures $h'-h_{1i}$ are nonnegative, and $1+h'\rho>0$, so that $\alpha_1'$ is defined. The rate ratio becomes $\exp(\alpha_1')=1+h'\rho$.

The likelihoods agree on their common parameter space. The fitted probabilities and AIC are unchanged if the maximizer over the larger space lies in that common space; separate expected counts under treatment and control may change.

\paragraph{Changing the assumed total exposure: log and logit links}\label{app:waking-curves}
Under the log link, $\pi_{ki}=\exp(\eta_{ki})$; under the logit link, $\pi_{ki}/(1-\pi_{ki})=\exp(\eta_{ki})$, with $\lambda_{ki}=\exp(\alpha_0+k\alpha_1+\bgamma^\top\bx_i)$, $k=0,1$, $\exp(\eta_{0i})=(h-h_{1i})\lambda_{0i}$ and $\exp(\eta_{1i})=h_{1i}\lambda_{1i}$. At a fixed covariate value, the probability of no event, $1-\pi_i=(1-\pi_{0i})(1-\pi_{1i})$, is then
\begin{align}
 1-\pi_i&=\{1-(h-h_{1i})\lambda_{0i}\}(1-h_{1i}\lambda_{1i})\qquad(\text{log})\nonumber\\
 (1-\pi_i)^{-1}&=\{1+(h-h_{1i})\lambda_{0i}\}(1+h_{1i}\lambda_{1i})\qquad(\text{logit})
 \label{eq:waking-factors}
\end{align}
Both expressions are quadratic in $h_{1i}$. Under the complementary log-log link, $-\log(1-\pi_i)=h\lambda_{0i}(1+h_{1i}\rho)$ is linear in $h_{1i}$ and depends on $h$ only through $h\lambda_{0i}$ and $\rho$.

For the log and logit links, suppose that two sets of coefficients, at totals $h$ and $h'$, give the same event probabilities for every treatment exposure in an interval of positive length. The corresponding quadratics in (\ref{eq:waking-factors}) must then have the same roots and coefficients. Under the log link, assume $h\lambda_{0i}<1$, so that the unique positive root is $1/\lambda_{1i}$. Under the logit link, the unique negative root is $-1/\lambda_{1i}$. In either case, this identifies $\lambda_{1i}$. The quadratic coefficient, $-\lambda_{0i}\lambda_{1i}$, then identifies $\lambda_{0i}$, and the constant term, $1-h\lambda_{0i}$ or $1+h\lambda_{0i}$, identifies $h$. Thus $h'=h$: under these conditions, changing the coefficients cannot preserve the entire exposure--probability curve at a different total exposure.

The two fits can still be close at the observed exposures. When $(h-h_{1i})\lambda_{0i}$ and $h_{1i}\lambda_{1i}$ are small, all three links give approximately
\[
 \pi_i\approx(h-h_{1i})\lambda_{0i}+h_{1i}\lambda_{1i}=h\lambda_{0i}(1+h_{1i}\rho),
\]
so when events are rare, the log and logit fits change little with $h$.

\subsection{Count outcome}\label{app:count-info}

\subsubsection{Expected information via the E-step}\label{app:count-info-estep}
The following derivation proves equation~(\ref{eq:info-count}) in Section~\ref{sec:counts-em} of the main paper.
\begin{eqnarray*}
I(\bbeta) &=& \var\{\E[s_c(\bbeta;\bW_0,\bW_1)|\bW]\}
\\&=& \var\left(
\sum_{i=1}^N W_i\bx_{Ci}\right)
\\&=&
\sum_{i=1}^N \mu_i\bx_{Ci}\bx_{Ci}^\top
\\&=& \bX_C^\top D_C \bX_C,
\end{eqnarray*}

\subsubsection{Score and expected information of the observed-data log-likelihood}\label{app:count-info-direct}
The following derivations give the score and expected information in Section~\ref{sec:counts-direct} of the main paper.
\begin{eqnarray*}
s(\bbeta;\bW) &=& \frac{\partial \ell(\bbeta;\bW)}{\partial \bbeta}
\\&=& \sum_{k,i}\frac{\partial \ell(\bbeta;\bW)}{\partial \mu_i}
\cdot \frac{\partial \mu_i}{\partial \mu_{ki}}
\cdot \frac{\partial \mu_{ki}}{\partial \eta_{ki}}
\cdot \frac{\partial \eta_{ki}}{\partial \bbeta}
\\&=&
\sum_{k,i}\frac{W_i-\mu_i}{\mu_i}
\cdot G'(\eta_{ki})\cdot \bx_{ki}
\\&=& \sum_{i=1}^N(W_i-\mu_i)\bx_{Ci}
\\&=& \bX_C^\top(\bW-\boldsymbol{\mu})
\end{eqnarray*}
where $\boldsymbol{\mu}=(\mu_1,\ldots, \mu_N)^\top$.

The expected information is
\begin{eqnarray*}
    I(\bbeta) &=& \E\left[-\frac{\partial s(\bbeta;\bW)}{\partial \bbeta^\top}\right]
    \\&=& \sum_{i=1}^N\bx_{Ci}\frac{\partial \mu_i}{\partial \bbeta^\top}
    \\&=& \sum_{i=1}^N \mu_i \bx_{Ci}\bx_{Ci}^\top
    \\&=& \bX_C^\top D_C \bX_C,
\end{eqnarray*}
which is identical to (\ref{eq:info-count}).

\subsubsection{Observed information via Louis' formula}\label{app:count-info-louis}
Given $W_i$, the count $W_{0i}$ is binomial with $W_i$ trials and probability $p_{0i}=\mu_{0i}/\mu_i$, and $W_{1i}=W_i-W_{0i}$. With $p_{1i}=\mu_{1i}/\mu_i$,
\[
 \Var(W_{0i}\mid W_i)=\Var(W_{1i}\mid W_i)=-\Cov(W_{0i},W_{1i}\mid W_i)=W_ip_{0i}p_{1i}.
\]
The conditional covariance of the complete-data score (\ref{eq:complete-score-count}) is therefore
\[
 \Var(s_c\mid\bW)=\sum_{i=1}^N W_ip_{0i}p_{1i}\,(w_{0i}\bx_{0i}-w_{1i}\bx_{1i})(w_{0i}\bx_{0i}-w_{1i}\bx_{1i})^\top.
\]
The complete-data observed information is $J_c=\sum_{k,i}\{\mu_{ki}w_{ki}^2-(W_{ki}-\mu_{ki})w'_{ki}\}\bx_{ki}\bx_{ki}^\top$, where $w'_{ki}=\partial w_{ki}/\partial\eta_{ki}$. Its conditional expectation replaces $W_{ki}$ by $\tW_{ki}=p_{ki}W_i$. Louis' formula, in the form (\ref{eq:louis}) of Section~\ref{app:louis}, gives
\[
 J(\bbeta;\bW)=\E(J_c\mid\bW)-\Var(s_c\mid\bW).
\]
For the log link, $w_{ki}=1$ and $w'_{ki}=0$, so
\[
 J(\bbeta;\bW)=\sum_{i=1}^N\left\{\mu_{0i}\bx_{0i}\bx_{0i}^\top+\mu_{1i}\bx_{1i}\bx_{1i}^\top
 -W_i\frac{\mu_{0i}\mu_{1i}}{\mu_i^2}(\bx_{1i}-\bx_{0i})(\bx_{1i}-\bx_{0i})^\top\right\}.
\]
Because $\bx_{1i}-\bx_{0i}=(0,1,\mathbf 0^\top)^\top$, the last term changes only the diagonal element for $\alpha_1$. Since $\E(W_i)=\mu_i$, $\E\{J(\bbeta;\bW)\}=I(\bbeta)$ in (\ref{eq:info-count}). The inverse $J(\hbbeta)^{-1}$ is an alternative estimator of $\Var(\hbbeta)$.

\section{Simulations}\label{app:simulation-checks}
\FloatBarrier
\subsection{Binary outcome}
Table~\ref{tab:sim_all72} gives EMG results under the complementary log-log link for the 72 settings of Section~\ref{sec:simulation_design} of the main paper. Performance measures use valid fits as defined there, with the same criteria applied separately to each link.
\begin{table}[!tbp]
\begin{singlespace}\centering\footnotesize\setlength{\tabcolsep}{2.5pt}
\caption{Binary simulation: estimation of $\alpha_1=\log R$ under the complementary log-log link in all 72 settings}\label{tab:sim_all72}
\begin{tabular}{@{}r@{\hspace{9pt}}r@{\hspace{6pt}}rrrrrr@{\hspace{10pt}}rrrrrr@{}}
\toprule
 & & \multicolumn{6}{c}{\boldmath$N=10{,}000$} & \multicolumn{6}{c}{\boldmath$N=1{,}000$} \\
\cmidrule(lr){3-8}\cmidrule(l){9-14}
\makecell[r]{$\boldsymbol{p}$\\\textbf{(\%)}} & \makecell[r]{\textbf{Rate}\\\textbf{ratio}} & \makecell[r]{\textbf{Exp.\ trt}\\\textbf{events}} & \makecell[r]{\textbf{Boundary}\\\textbf{(\%)}} & \makecell[r]{\textbf{Valid}\\\textbf{fits}} & \textbf{Bias} & \textbf{RMSE} & \makecell[r]{\textbf{Cov.}\\\textbf{(\%)}} & \makecell[r]{\textbf{Exp.\ trt}\\\textbf{events}} & \makecell[r]{\textbf{Boundary}\\\textbf{(\%)}} & \makecell[r]{\textbf{Valid}\\\textbf{fits}} & \textbf{Bias} & \textbf{RMSE} & \makecell[r]{\textbf{Cov.}\\\textbf{(\%)}} \\
\midrule
\multicolumn{14}{@{}l}{\fontsize{8.5}{10}\selectfont\itshape{Short exposure, lower variability: $\mathrm{Beta}(1,30)$; mean 3.23\%, SD 3.12\%}} \\
\addlinespace[2pt]
1 & 0.5 & 1.40 & 44.84 & 5,494 & 1.341 & 1.752 & 71.0 & 0.14 & 53.61 & 4,634 & 2.843 & 3.164 & 62.8 \\
 & 1 & 2.76 & 38.56 & 6,123 & 0.778 & 1.361 & 79.5 & 0.28 & 52.43 & 4,751 & 2.157 & 2.576 & 73.0 \\
 & 2 & 5.36 & 27.60 & 7,222 & 0.285 & 1.094 & 87.2 & 0.54 & 49.01 & 5,089 & 1.556 & 2.083 & 82.3 \\
 & 7 & 16.57 & 4.67 & 9,531 & $-$0.117 & 0.828 & 95.0 & 1.66 & 35.83 & 6,408 & 0.561 & 1.473 & 97.1 \\
\addlinespace[2pt]
5 & 0.5 & 7.13 & 36.83 & 6,276 & 0.629 & 1.178 & 79.6 & 0.71 & 47.63 & 5,224 & 1.760 & 2.100 & 66.2 \\
 & 1 & 14.06 & 23.81 & 7,590 & 0.164 & 0.978 & 87.2 & 1.41 & 43.70 & 5,617 & 1.106 & 1.592 & 76.1 \\
 & 2 & 27.38 & 9.10 & 9,071 & $-$0.109 & 0.858 & 92.7 & 2.74 & 34.90 & 6,497 & 0.578 & 1.278 & 83.8 \\
 & 7 & 84.60 & 0.00 & 10,000 & $-$0.033 & 0.345 & 95.9 & 8.46 & 11.09 & 8,884 & $-$0.043 & 0.995 & 93.9 \\
\addlinespace[2pt]
15 & 0.5 & 22.59 & 25.85 & 7,354 & 0.246 & 0.963 & 85.2 & 2.26 & 42.66 & 5,707 & 1.148 & 1.578 & 73.3 \\
 & 1 & 44.56 & 10.47 & 8,919 & $-$0.084 & 0.856 & 91.2 & 4.46 & 36.15 & 6,359 & 0.616 & 1.246 & 80.3 \\
 & 2 & 86.75 & 0.80 & 9,915 & $-$0.101 & 0.577 & 94.3 & 8.68 & 23.00 & 7,687 & 0.132 & 1.041 & 88.2 \\
 & 7 & 268.41 & 0.00 & 10,000 & $-$0.011 & 0.187 & 95.3 & 26.84 & 1.59 & 9,840 & $-$0.101 & 0.698 & 95.2 \\
\addlinespace[5pt]
\multicolumn{14}{@{}l}{\fontsize{8.5}{10}\selectfont\itshape{Short exposure, higher variability: $\mathrm{Beta}(0.1,3)$; mean 3.23\%, SD 8.73\%}} \\
\addlinespace[2pt]
1 & 0.5 & 1.40 & 37.47 & 6,251 & 0.440 & 1.177 & 85.4 & 0.14 & 61.67 & 3,833 & 1.952 & 2.258 & 68.8 \\
 & 1 & 2.76 & 23.94 & 7,600 & 0.054 & 0.998 & 91.2 & 0.28 & 56.78 & 4,321 & 1.339 & 1.741 & 79.1 \\
 & 2 & 5.36 & 8.30 & 9,168 & $-$0.134 & 0.877 & 94.6 & 0.54 & 48.49 & 5,147 & 0.823 & 1.368 & 86.4 \\
 & 7 & 16.58 & 0.04 & 9,996 & $-$0.054 & 0.415 & 96.8 & 1.66 & 22.53 & 7,747 & 0.138 & 1.031 & 95.2 \\
\addlinespace[2pt]
5 & 0.5 & 7.13 & 15.17 & 8,477 & $-$0.068 & 0.951 & 91.4 & 0.71 & 45.14 & 5,482 & 0.803 & 1.360 & 82.1 \\
 & 1 & 14.06 & 2.89 & 9,709 & $-$0.139 & 0.722 & 94.5 & 1.41 & 34.12 & 6,580 & 0.317 & 1.121 & 88.0 \\
 & 2 & 27.38 & 0.03 & 9,997 & $-$0.056 & 0.395 & 95.7 & 2.74 & 18.93 & 8,106 & $-$0.004 & 0.998 & 92.7 \\
 & 7 & 84.91 & 0.00 & 10,000 & $-$0.007 & 0.170 & 95.1 & 8.49 & 0.78 & 9,922 & $-$0.102 & 0.645 & 95.8 \\
\addlinespace[2pt]
15 & 0.5 & 22.59 & 2.88 & 9,710 & $-$0.132 & 0.716 & 94.1 & 2.26 & 32.24 & 6,772 & 0.279 & 1.100 & 87.4 \\
 & 1 & 44.56 & 0.03 & 9,997 & $-$0.048 & 0.375 & 95.6 & 4.46 & 16.04 & 8,393 & $-$0.047 & 0.968 & 91.9 \\
 & 2 & 86.79 & 0.00 & 10,000 & $-$0.018 & 0.204 & 95.6 & 8.68 & 3.87 & 9,612 & $-$0.124 & 0.766 & 94.7 \\
 & 7 & 271.40 & 0.00 & 10,000 & $-$0.002 & 0.099 & 95.3 & 27.14 & 0.00 & 10,000 & $-$0.023 & 0.333 & 96.1 \\
\addlinespace[5pt]
\multicolumn{14}{@{}l}{\fontsize{8.5}{10}\selectfont\itshape{Longer exposure: $\mathrm{Beta}(2,5)$; mean 28.57\%, SD 15.97\%}} \\
\addlinespace[2pt]
1 & 0.5 & 13.89 & 8.58 & 9,139 & $-$0.095 & 0.923 & 93.7 & 1.39 & 40.74 & 5,919 & 0.770 & 1.614 & 95.8 \\
 & 1 & 24.41 & 0.83 & 9,916 & $-$0.089 & 0.682 & 96.2 & 2.44 & 32.68 & 6,725 & 0.354 & 1.413 & 99.0 \\
 & 2 & 39.28 & 0.00 & 9,999 & $-$0.030 & 0.486 & 97.1 & 3.93 & 25.17 & 7,479 & 0.058 & 1.341 & 100.0 \\
 & 7 & 69.60 & 0.10 & 9,990 & 0.045 & 0.464 & 96.9 & 6.96 & 33.57 & 6,636 & $-$0.391 & 1.325 & 100.0 \\
\addlinespace[2pt]
5 & 0.5 & 70.91 & 0.08 & 9,991 & $-$0.065 & 0.442 & 95.4 & 7.09 & 17.61 & 8,230 & 0.057 & 1.047 & 92.6 \\
 & 1 & 124.57 & 0.00 & 10,000 & $-$0.017 & 0.257 & 96.1 & 12.46 & 5.03 & 9,495 & $-$0.093 & 0.935 & 95.2 \\
 & 2 & 200.56 & 0.00 & 10,000 & $-$0.006 & 0.197 & 95.5 & 20.06 & 0.54 & 9,946 & $-$0.041 & 0.730 & 97.3 \\
 & 7 & 356.63 & 0.00 & 10,000 & 0.008 & 0.193 & 95.3 & 35.66 & 2.13 & 9,786 & 0.073 & 0.687 & 98.4 \\
\addlinespace[2pt]
15 & 0.5 & 224.81 & 0.00 & 10,000 & $-$0.017 & 0.222 & 95.6 & 22.48 & 4.28 & 9,568 & $-$0.124 & 0.818 & 94.4 \\
 & 1 & 394.69 & 0.00 & 10,000 & $-$0.004 & 0.144 & 95.2 & 39.47 & 0.16 & 9,984 & $-$0.062 & 0.530 & 96.3 \\
 & 2 & 636.20 & 0.00 & 10,000 & $-$0.004 & 0.113 & 95.2 & 63.62 & 0.00 & 10,000 & $-$0.008 & 0.369 & 96.7 \\
 & 7 & 1142.23 & 0.00 & 10,000 & 0.002 & 0.110 & 94.9 & 114.22 & 0.01 & 9,999 & 0.024 & 0.363 & 95.8 \\
\bottomrule
\end{tabular}
\par\vspace{4pt}{\footnotesize\raggedright\setstretch{1.1}Each setting comprises 10,000 simulated datasets. Here $p$ is the marginal event probability (reported as a percentage), so the expected number of subjects with an event is $Np$. For each combination of $p$, rate ratio and exposure distribution,
$\alpha_0$ is determined by numerical integration and root finding so that the event probability, averaged over the exposure distribution, equals $p$. Total exposure is $h_i=1$, with $\Pr(h_{1i}>0)=0.85$. The reported exposure means and SDs describe the proportion of total exposure under treatment, $h_{1i}$, conditional on $h_{1i}>0$. Exp.\ trt events is the expected latent event count during treatment, $N\lambda_1 E(h_{1i})$, where the expectation includes subjects with zero treatment exposure. Boundary is the percentage of all 10,000 datasets with $\hat\lambda_1=0$ or $\hat\lambda_0=0$. Bias and RMSE (root mean squared error) summarize estimation of
$\alpha_1=\log R$; Cov.\ is the percentage of 95\% Wald intervals
containing $\alpha_1$. These three measures are computed using only the valid fits. The largest Monte Carlo standard errors across settings are approximately 0.02 for Bias, 0.5 percentage points for Boundary, and 0.8 percentage points for Cov.\par}
\end{singlespace}
\end{table}

\subsubsection{Comparison of $I(\hbbeta)^{-1}$ and $J(\hbbeta)^{-1}$ as variance estimators}\label{app:info-comparison}
Table~\ref{tab:information_comparison} compares standard errors and Wald coverage for $\alpha_1$ using observed and expected information at the same EMG estimates. The two approaches gave similar results in the 37 settings with at least 10 expected events during treatment exposure: median standard errors differed by at most 1.2\% and coverage by at most 0.3 percentage points. With fewer events, expected information gave mean standard errors up to 31\% smaller and coverage up to 3.7 percentage points lower, in the context of poor statistical performance:
For example, under $\mathrm{Beta}(1,30)$ exposure with $N=1{,}000$, $p=1\%$ and $R=1$, coverage was  73.0\% with observed information and 69.3\% with expected information.

\begin{table}[!htbp]
\begin{singlespace}\centering\footnotesize\setlength{\tabcolsep}{3pt}
\caption{Binary simulation: Wald coverage and standard errors of $\hat\alpha_1$ with observed and expected information in all 72 settings}\label{tab:information_comparison}
\begin{tabular}{@{}r@{\hspace{8pt}}r@{\hspace{6pt}}rrrrrr@{\hspace{10pt}}rrrrrr@{}}
\toprule
 & & \multicolumn{6}{c}{\boldmath$N=10{,}000$} & \multicolumn{6}{c}{\boldmath$N=1{,}000$} \\
\cmidrule(lr){3-8}\cmidrule(l){9-14}
 & & \multicolumn{2}{c}{\textbf{Cov.\ (\%)}} & \multicolumn{2}{c}{\textbf{Mean SE}} & \multicolumn{2}{c}{\textbf{Median SE}} & \multicolumn{2}{c}{\textbf{Cov.\ (\%)}} & \multicolumn{2}{c}{\textbf{Mean SE}} & \multicolumn{2}{c}{\textbf{Median SE}} \\
\cmidrule(lr){3-4}\cmidrule(lr){5-6}\cmidrule(lr){7-8}\cmidrule(lr){9-10}\cmidrule(lr){11-12}\cmidrule(l){13-14}
\makecell[r]{$\boldsymbol{p}$\\\textbf{(\%)}} & $\boldsymbol{R}$ & \textbf{Obs.} & \textbf{Exp.} & \textbf{Obs.} & \textbf{Exp.} & \textbf{Obs.} & \textbf{Exp.} & \textbf{Obs.} & \textbf{Exp.} & \textbf{Obs.} & \textbf{Exp.} & \textbf{Obs.} & \textbf{Exp.} \\
\midrule
\multicolumn{14}{@{}l}{\fontsize{8.5}{10}\selectfont\itshape{Short exposure, lower variability: $\mathrm{Beta}(1,30)$}} \\
\addlinespace[2pt]
1 & 0.5 & 71.0 & 71.0 & 4.309 & 4.300 & 1.519 & 1.512 & 62.8 & 60.0 & 5.279 & 4.834 & 2.054 & 1.951 \\
 & 1 & 79.5 & 79.1 & 3.877 & 3.879 & 1.357 & 1.362 & 73.0 & 69.3 & 5.695 & 5.439 & 2.014 & 1.912 \\
 & 2 & 87.2 & 87.0 & 3.042 & 3.039 & 1.177 & 1.176 & 82.3 & 79.5 & 5.067 & 4.854 & 1.917 & 1.857 \\
 & 7 & 95.0 & 94.9 & 1.182 & 1.169 & 0.677 & 0.676 & 97.1 & 97.0 & 4.100 & 3.760 & 1.703 & 1.658 \\
\addlinespace[2pt]
5 & 0.5 & 79.6 & 79.8 & 2.970 & 2.966 & 1.285 & 1.282 & 66.2 & 66.2 & 4.498 & 4.440 & 1.603 & 1.583 \\
 & 1 & 87.2 & 87.1 & 2.408 & 2.409 & 1.041 & 1.041 & 76.1 & 75.7 & 4.014 & 4.049 & 1.521 & 1.519 \\
 & 2 & 92.7 & 92.7 & 1.488 & 1.487 & 0.733 & 0.736 & 83.8 & 83.3 & 3.714 & 3.694 & 1.347 & 1.347 \\
 & 7 & 95.9 & 95.9 & 0.337 & 0.336 & 0.309 & 0.309 & 93.9 & 93.8 & 1.922 & 1.884 & 0.914 & 0.907 \\
\addlinespace[2pt]
15 & 0.5 & 85.2 & 85.2 & 2.269 & 2.267 & 1.045 & 1.045 & 73.3 & 73.3 & 3.807 & 3.810 & 1.480 & 1.483 \\
 & 1 & 91.2 & 91.1 & 1.529 & 1.528 & 0.757 & 0.757 & 80.3 & 80.0 & 3.498 & 3.489 & 1.267 & 1.264 \\
 & 2 & 94.3 & 94.4 & 0.619 & 0.619 & 0.451 & 0.451 & 88.2 & 88.2 & 2.757 & 2.744 & 1.087 & 1.088 \\
 & 7 & 95.3 & 95.2 & 0.183 & 0.183 & 0.179 & 0.179 & 95.2 & 95.2 & 0.835 & 0.830 & 0.564 & 0.564 \\
\addlinespace[4pt]
\multicolumn{14}{@{}l}{\fontsize{8.5}{10}\selectfont\itshape{Short exposure, higher variability: $\mathrm{Beta}(0.1,3)$}} \\
\addlinespace[2pt]
1 & 0.5 & 85.4 & 85.2 & 4.140 & 3.721 & 1.265 & 1.263 & 68.8 & 67.3 & 4.491 & 3.108 & 1.602 & 1.578 \\
 & 1 & 91.2 & 91.3 & 2.670 & 2.414 & 1.023 & 1.029 & 79.1 & 78.0 & 3.851 & 2.826 & 1.506 & 1.522 \\
 & 2 & 94.6 & 94.5 & 1.540 & 1.397 & 0.731 & 0.733 & 86.4 & 86.1 & 3.265 & 2.461 & 1.352 & 1.380 \\
 & 7 & 96.8 & 96.8 & 0.396 & 0.394 & 0.369 & 0.368 & 95.2 & 94.9 & 1.944 & 1.591 & 1.081 & 1.078 \\
\addlinespace[2pt]
5 & 0.5 & 91.4 & 91.2 & 2.218 & 2.118 & 0.847 & 0.846 & 82.1 & 81.8 & 4.240 & 3.696 & 1.388 & 1.391 \\
 & 1 & 94.5 & 94.5 & 0.958 & 0.922 & 0.550 & 0.550 & 88.0 & 87.9 & 3.395 & 2.977 & 1.192 & 1.213 \\
 & 2 & 95.7 & 95.7 & 0.377 & 0.373 & 0.337 & 0.337 & 92.7 & 92.6 & 2.317 & 2.057 & 0.947 & 0.950 \\
 & 7 & 95.1 & 95.1 & 0.167 & 0.167 & 0.166 & 0.166 & 95.8 & 95.8 & 0.680 & 0.652 & 0.528 & 0.528 \\
\addlinespace[2pt]
15 & 0.5 & 94.1 & 94.0 & 0.945 & 0.927 & 0.540 & 0.541 & 87.4 & 87.3 & 3.475 & 3.277 & 1.178 & 1.178 \\
 & 1 & 95.6 & 95.6 & 0.354 & 0.353 & 0.317 & 0.317 & 91.9 & 91.7 & 2.172 & 2.042 & 0.904 & 0.913 \\
 & 2 & 95.6 & 95.7 & 0.201 & 0.201 & 0.197 & 0.197 & 94.7 & 94.7 & 1.038 & 0.982 & 0.616 & 0.618 \\
 & 7 & 95.3 & 95.2 & 0.100 & 0.100 & 0.099 & 0.099 & 96.1 & 96.2 & 0.327 & 0.327 & 0.314 & 0.315 \\
\addlinespace[4pt]
\multicolumn{14}{@{}l}{\fontsize{8.5}{10}\selectfont\itshape{Longer exposure: $\mathrm{Beta}(2,5)$}} \\
\addlinespace[2pt]
1 & 0.5 & 93.7 & 93.5 & 1.661 & 1.619 & 0.790 & 0.789 & 95.8 & 95.8 & 4.872 & 4.287 & 1.838 & 1.756 \\
 & 1 & 96.2 & 96.0 & 0.750 & 0.738 & 0.560 & 0.560 & 99.0 & 99.6 & 3.926 & 3.412 & 1.680 & 1.635 \\
 & 2 & 97.1 & 96.9 & 0.469 & 0.466 & 0.443 & 0.440 & 100.0 & 100.0 & 3.231 & 2.685 & 1.528 & 1.506 \\
 & 7 & 96.9 & 96.6 & 0.452 & 0.447 & 0.434 & 0.433 & 100.0 & 100.0 & 2.722 & 1.929 & 1.378 & 1.434 \\
\addlinespace[2pt]
5 & 0.5 & 95.4 & 95.4 & 0.423 & 0.422 & 0.371 & 0.372 & 92.6 & 92.5 & 2.312 & 2.237 & 1.016 & 1.013 \\
 & 1 & 96.1 & 96.1 & 0.257 & 0.257 & 0.250 & 0.250 & 95.2 & 95.0 & 1.379 & 1.336 & 0.783 & 0.780 \\
 & 2 & 95.5 & 95.4 & 0.197 & 0.197 & 0.196 & 0.195 & 97.3 & 97.0 & 0.753 & 0.739 & 0.638 & 0.631 \\
 & 7 & 95.3 & 95.4 & 0.192 & 0.192 & 0.191 & 0.191 & 98.4 & 98.3 & 0.728 & 0.666 & 0.625 & 0.621 \\
\addlinespace[2pt]
15 & 0.5 & 95.6 & 95.7 & 0.220 & 0.220 & 0.214 & 0.214 & 94.4 & 94.4 & 1.171 & 1.158 & 0.664 & 0.667 \\
 & 1 & 95.2 & 95.1 & 0.145 & 0.145 & 0.144 & 0.144 & 96.3 & 96.2 & 0.537 & 0.533 & 0.458 & 0.458 \\
 & 2 & 95.2 & 95.1 & 0.113 & 0.113 & 0.113 & 0.113 & 96.7 & 96.6 & 0.369 & 0.369 & 0.359 & 0.358 \\
 & 7 & 94.9 & 94.8 & 0.109 & 0.109 & 0.109 & 0.109 & 95.8 & 95.9 & 0.355 & 0.353 & 0.348 & 0.347 \\
\bottomrule
\end{tabular}
\par\vspace{4pt}{\footnotesize\raggedright\setstretch{1.1}Each setting has 10,000 datasets, generated as described in Section~\ref{sec:simulation_design} of the main paper; Table~\ref{tab:sim_all72} gives the expected numbers of events during treatment exposure. Both information matrices are evaluated at the same EMG estimates of the valid fits (Table~\ref{tab:sim_all72}); both could be inverted in every valid fit. Obs., observed information; Exp., expected information; Cov., coverage of the 95\% Wald interval; SE, standard error of $\hat\alpha_1$. The main results use observed information.\par}
\end{singlespace}
\end{table}

\subsubsection{Different links}\label{app:sim-design}
Table~\ref{tab:sim_links72} gives direct Fisher scoring results under the logit and log links for the same datasets, generated under the complementary log-log model. Boundary maxima were classified separately under each link. For the log link, infeasible starting values were adjusted by lowering $\alpha_0$ until the largest linear predictor was $-0.1$.

The three links gave similar estimates with few events during treatment exposure. Under the logit and log links, however, $\alpha_1$ is the log odds ratio and log risk ratio for unit exposures, respectively; the reported bias and coverage relative to the generating $\log R$ reflect this difference.

\begin{table}[!htbp]
\begin{singlespace}\centering\footnotesize\setlength{\tabcolsep}{2.5pt}
\caption{Binary simulation: estimates of $\alpha_1$ under the logit and log links, compared with the generating $\log R$, in all 72 settings}\label{tab:sim_links72}
\begin{tabular}{@{}r@{\hspace{9pt}}rrrr@{\hspace{6pt}}rrr@{\hspace{10pt}}rrr@{\hspace{6pt}}rrr@{}}
\toprule
 & & \multicolumn{6}{c}{\boldmath$N=10{,}000$} & \multicolumn{6}{c}{\boldmath$N=1{,}000$} \\
\cmidrule(lr){3-8}\cmidrule(l){9-14}
 & & \multicolumn{3}{c}{\textbf{Logit}} & \multicolumn{3}{c}{\textbf{Log}} & \multicolumn{3}{c}{\textbf{Logit}} & \multicolumn{3}{c}{\textbf{Log}} \\
\cmidrule(lr){3-5}\cmidrule(lr){6-8}\cmidrule(lr){9-11}\cmidrule(l){12-14}
\makecell[r]{$\boldsymbol{p}$\\\textbf{(\%)}} & $\boldsymbol{R}$ & \textbf{Bias} & \textbf{RMSE} & \makecell[r]{\textbf{Cov.}\\\textbf{(\%)}} & \textbf{Bias} & \textbf{RMSE} & \makecell[r]{\textbf{Cov.}\\\textbf{(\%)}} & \textbf{Bias} & \textbf{RMSE} & \makecell[r]{\textbf{Cov.}\\\textbf{(\%)}} & \textbf{Bias} & \textbf{RMSE} & \makecell[r]{\textbf{Cov.}\\\textbf{(\%)}} \\
\midrule
\multicolumn{14}{@{}l}{\fontsize{8.5}{10}\selectfont\itshape{Short exposure, lower variability: $\mathrm{Beta}(1,30)$}} \\
\addlinespace[2pt]
1 & 0.5 & 1.318 & 1.768 & 71.3 & 1.327 & 1.768 & 71.0 & 2.854 & 3.171 & 62.7 & 2.852 & 3.168 & 62.3 \\
 & 1 & 0.759 & 1.391 & 79.6 & 0.764 & 1.390 & 79.3 & 2.182 & 2.613 & 72.7 & 2.176 & 2.593 & 72.3 \\
 & 2 & 0.266 & 1.135 & 87.5 & 0.276 & 1.119 & 87.1 & 1.572 & 2.110 & 82.5 & 1.554 & 2.096 & 81.8 \\
 & 7 & $-$0.120 & 0.839 & 95.1 & $-$0.118 & 0.833 & 95.0 & 0.573 & 1.482 & 97.1 & 0.561 & 1.471 & 97.0 \\
\addlinespace[2pt]
5 & 0.5 & 0.566 & 1.221 & 80.9 & 0.626 & 1.244 & 79.0 & 1.738 & 2.109 & 67.0 & 1.763 & 2.111 & 65.5 \\
 & 1 & 0.111 & 1.041 & 87.8 & 0.175 & 1.024 & 86.5 & 1.081 & 1.609 & 76.8 & 1.110 & 1.605 & 75.2 \\
 & 2 & $-$0.149 & 0.916 & 93.4 & $-$0.094 & 0.886 & 92.0 & 0.562 & 1.304 & 84.4 & 0.569 & 1.309 & 83.1 \\
 & 7 & $-$0.041 & 0.348 & 96.1 & $-$0.025 & 0.336 & 95.7 & $-$0.049 & 1.027 & 94.2 & $-$0.049 & 1.008 & 93.6 \\
\addlinespace[2pt]
15 & 0.5 & 0.079 & 1.062 & 88.7 & 0.322 & 1.072 & 81.8 & 1.051 & 1.575 & 75.8 & 1.205 & 1.633 & 71.0 \\
 & 1 & $-$0.229 & 0.970 & 93.7 & 0.019 & 0.886 & 87.4 & 0.549 & 1.244 & 82.9 & 0.648 & 1.305 & 78.1 \\
 & 2 & $-$0.208 & 0.650 & 96.6 & $-$0.003 & 0.556 & 91.2 & 0.068 & 1.073 & 89.9 & 0.177 & 1.066 & 86.2 \\
 & 7 & $-$0.040 & 0.202 & 95.7 & 0.013 & 0.175 & 94.5 & $-$0.130 & 0.748 & 95.8 & $-$0.081 & 0.678 & 94.4 \\
\addlinespace[4pt]
\multicolumn{14}{@{}l}{\fontsize{8.5}{10}\selectfont\itshape{Short exposure, higher variability: $\mathrm{Beta}(0.1,3)$}} \\
\addlinespace[2pt]
1 & 0.5 & 0.444 & 1.193 & 85.4 & 0.448 & 1.198 & 85.1 & 1.956 & 2.271 & 69.0 & 1.944 & 2.250 & 68.6 \\
 & 1 & 0.060 & 1.002 & 91.3 & 0.065 & 0.998 & 91.0 & 1.344 & 1.755 & 79.3 & 1.334 & 1.730 & 78.8 \\
 & 2 & $-$0.132 & 0.880 & 94.6 & $-$0.129 & 0.875 & 94.5 & 0.830 & 1.384 & 86.6 & 0.813 & 1.366 & 86.2 \\
 & 7 & $-$0.048 & 0.418 & 96.6 & $-$0.061 & 0.412 & 96.8 & 0.149 & 1.040 & 95.0 & 0.126 & 1.021 & 95.2 \\
\addlinespace[2pt]
5 & 0.5 & $-$0.102 & 0.971 & 92.3 & $-$0.027 & 0.936 & 90.3 & 0.805 & 1.371 & 82.5 & 0.813 & 1.375 & 81.3 \\
 & 1 & $-$0.167 & 0.737 & 95.2 & $-$0.111 & 0.716 & 93.9 & 0.320 & 1.142 & 88.4 & 0.320 & 1.143 & 87.2 \\
 & 2 & $-$0.069 & 0.407 & 96.2 & $-$0.043 & 0.384 & 95.2 & 0.001 & 1.032 & 92.9 & $-$0.001 & 0.991 & 92.3 \\
 & 7 & 0.022 & 0.178 & 94.5 & $-$0.040 & 0.166 & 95.2 & $-$0.071 & 0.662 & 95.5 & $-$0.136 & 0.628 & 96.2 \\
\addlinespace[2pt]
15 & 0.5 & $-$0.298 & 0.830 & 96.8 & 0.023 & 0.648 & 89.7 & 0.216 & 1.114 & 90.0 & 0.355 & 1.102 & 84.6 \\
 & 1 & $-$0.150 & 0.434 & 97.8 & 0.045 & 0.345 & 91.5 & $-$0.108 & 1.018 & 93.5 & 0.020 & 0.937 & 89.8 \\
 & 2 & $-$0.063 & 0.228 & 96.4 & 0.018 & 0.187 & 94.1 & $-$0.163 & 0.823 & 95.6 & $-$0.095 & 0.728 & 93.5 \\
 & 7 & 0.074 & 0.134 & 89.3 & $-$0.124 & 0.149 & 69.9 & 0.054 & 0.375 & 94.4 & $-$0.151 & 0.315 & 97.3 \\
\addlinespace[4pt]
\multicolumn{14}{@{}l}{\fontsize{8.5}{10}\selectfont\itshape{Longer exposure: $\mathrm{Beta}(2,5)$}} \\
\addlinespace[2pt]
1 & 0.5 & $-$0.106 & 0.949 & 93.9 & $-$0.091 & 0.938 & 93.6 & 0.783 & 1.618 & 95.8 & 0.784 & 1.619 & 95.6 \\
 & 1 & $-$0.094 & 0.689 & 96.3 & $-$0.085 & 0.680 & 96.2 & 0.356 & 1.409 & 99.1 & 0.355 & 1.399 & 99.1 \\
 & 2 & $-$0.030 & 0.489 & 97.1 & $-$0.030 & 0.487 & 97.1 & 0.054 & 1.332 & 100.0 & 0.052 & 1.323 & 100.0 \\
 & 7 & 0.052 & 0.468 & 97.0 & 0.038 & 0.461 & 96.9 & $-$0.414 & 1.324 & 100.0 & $-$0.422 & 1.311 & 100.0 \\
\addlinespace[2pt]
5 & 0.5 & $-$0.117 & 0.475 & 96.5 & $-$0.016 & 0.418 & 93.9 & 0.036 & 1.064 & 93.2 & 0.081 & 1.049 & 92.1 \\
 & 1 & $-$0.040 & 0.268 & 96.7 & 0.006 & 0.249 & 95.4 & $-$0.116 & 0.975 & 95.4 & $-$0.074 & 0.919 & 94.8 \\
 & 2 & $-$0.008 & 0.203 & 95.5 & $-$0.005 & 0.191 & 95.5 & $-$0.044 & 0.761 & 97.4 & $-$0.042 & 0.713 & 97.3 \\
 & 7 & 0.042 & 0.202 & 95.1 & $-$0.028 & 0.189 & 94.8 & 0.102 & 0.690 & 98.6 & 0.034 & 0.655 & 98.2 \\
\addlinespace[2pt]
15 & 0.5 & $-$0.173 & 0.307 & 96.2 & 0.125 & 0.231 & 85.3 & $-$0.272 & 0.933 & 96.6 & 0.010 & 0.763 & 91.2 \\
 & 1 & $-$0.078 & 0.177 & 94.2 & 0.063 & 0.145 & 91.0 & $-$0.142 & 0.597 & 97.4 & 0.012 & 0.471 & 94.8 \\
 & 2 & $-$0.011 & 0.124 & 95.2 & $-$0.002 & 0.103 & 95.2 & $-$0.015 & 0.405 & 96.9 & $-$0.006 & 0.335 & 96.7 \\
 & 7 & 0.105 & 0.159 & 86.0 & $-$0.112 & 0.150 & 78.7 & 0.130 & 0.412 & 95.4 & $-$0.093 & 0.343 & 93.6 \\
\bottomrule
\end{tabular}
\par\vspace{4pt}{\footnotesize\raggedright\setstretch{1.1}The datasets are those of Table~\ref{tab:sim_all72}, generated under the complementary log-log model. Fits use direct Fisher scoring with step halving. Boundary maxima were classified separately under each link; the boundary percentages differed from those under the complementary log-log link by at most 3.4 percentage points. Valid fits were 99.7\% (logit) and 99.7\% (log) of the remaining datasets. Bias, RMSE and Cov.\ (coverage of the 95\% Wald interval) compare $\hat\alpha_1$ with the generating complementary log-log value $\log R$; under the logit and log links, $\alpha_1$ is the log odds ratio and log risk ratio for unit exposures, a different quantity, so these measures need not approach their nominal values as information increases.\par}
\end{singlespace}
\end{table}

\subsubsection{Boundary datasets}\label{app:sim-boundary}
For the complementary log-log model without covariates, the log-likelihood is concave in $(\lambda_0,\lambda_1)$:
\[
 \ell(\lambda_0,\lambda_1)
 =\sum_{i:\,Y_i=1}\log\{1-\exp(-\mu_i)\}
  -\sum_{i:\,Y_i=0}\mu_i,
 \qquad \mu_i=\lambda_0h_{0i}+\lambda_1h_{1i}.
\]
Let $\tilde\lambda_0$ maximize $\ell(\lambda_0,0)$ and put $\tilde\pi_i=1-\exp(-\tilde\lambda_0h_{0i})$. The maximum lies on this edge if
\[
 \left.\frac{\partial\ell}{\partial\lambda_1}\right|_{(\tilde\lambda_0,0)}
 =\sum_i h_{1i}\left(\frac{Y_i}{\tilde\pi_i}-1\right)\leq0.
\]
Interchanging control and treatment gives the check for $\lambda_0=0$; edges assigning zero probability to an observed event are excluded. Six datasets had no events and were classified separately from boundary datasets; they are excluded from boundary percentages and valid-fit summaries.

\FloatBarrier
\subsubsection{Computation time}\label{app:timing}
Table~\ref{tab:timing_common} gives computation times for Section~\ref{sec:simulation_computation} of the main paper. All algorithms used the same starting values and stopping rule: $J$ positive definite and $s^\top J^{-1}s/2<10^{-8}$, checked after every update \citep{Proust2017lcmm}.
\begin{table}[!tbp]
\begin{singlespace}\centering\footnotesize\setlength{\tabcolsep}{5pt}
\caption{Computation time at a common accuracy, 250 datasets with finite MLEs in each of four settings per model.}\label{tab:timing_common}
\begin{tabular}{@{}lccc@{}}
\toprule
Algorithm & Median time (IQR), s & \makecell{Median time\\ratio} & \makecell{Range over\\settings} \\
\midrule
\multicolumn{4}{@{}l}{\textit{A. Binary model}} \\
Direct Fisher scoring & 0.013 (0.012--0.015) & 1 & -- \\
EM, Newton--Raphson M-step & 0.141 (0.089--0.392) & 10.1 & 4.8--34.9 \\
EM, Fisher scoring M-step & 0.142 (0.090--0.397) & 10.2 & 4.7--35.6 \\
EMG, one Newton--Raphson step & 0.113 (0.067--0.304) & 8.0 & 3.6--26.8 \\
EMG, one Fisher scoring step & 0.109 (0.066--0.301) & 7.7 & 3.6--26.6 \\
EM $\to$ Fisher scoring & 0.091 (0.082--0.104) & 7.0 & 4.4--7.9 \\
EMG $\to$ Fisher scoring & 0.063 (0.058--0.068) & 4.7 & 3.3--5.3 \\
\addlinespace[4pt]
\multicolumn{4}{@{}l}{\textit{B. Count model}} \\
Direct Fisher scoring & 0.011 (0.009--0.012) & 1 & -- \\
EM, iterated M-step & 0.108 (0.099--0.115) & 10.3 & 9.9--11.1 \\
EM, closed-form M-step & 0.066 (0.061--0.072) & 6.4 & 6.2--6.9 \\
EMG & 0.086 (0.079--0.092) & 8.1 & 7.9--8.9 \\
EM $\to$ Fisher scoring & 0.060 (0.058--0.062) & 5.5 & 5.3--6.8 \\
EMG $\to$ Fisher scoring & 0.077 (0.074--0.081) & 7.2 & 6.7--8.7 \\
\bottomrule
\end{tabular}
\par\vspace{4pt}{\footnotesize\raggedright\setstretch{1.15}Median time ratio: median over datasets of the time divided by that of direct Fisher scoring on the same dataset; range over settings: smallest and largest of the four setting-specific medians. Binary settings: $N=10{,}000$, $R=2$, $p=1\%$ or 15\%, $\mathrm{Beta}(0.1,3)$ or $\mathrm{Beta}(1,30)$ treatment exposure; count settings: $N=10{,}000$, $R=2$, $\mathrm{Beta}(2,2)$ treatment exposure for the 85\% of subjects with treatment exposure, and event rate under control 0.1, 0.5, 1 or 3. Every algorithm converged in all 1,000 datasets of each model, to within $10^{-7}$ of an independently computed maximum log-likelihood. Times include the stopping checks and the final covariance matrix. Hybrids ($\to$) use at most 25 EM or EMG iterations, then direct Fisher scoring. All algorithms had a limit of 20,000 updates (direct Fisher scoring in the 72-setting simulation: 1,000). EM, expectation--maximization; EMG, EM gradient; IQR, interquartile range.\par}
\end{singlespace}
\end{table}

\FloatBarrier
\subsection{Count outcome}
\FloatBarrier
\subsubsection{Count model in the 72 settings}\label{app:count-72}
Table~\ref{tab:sim_count_all72} gives count model results for the 72 count settings of Section~\ref{sec:simulation_counts} of the main paper. For the Poisson model without covariates, the maximum on each edge has a closed form. With $T=\sum_iW_i>0$, $H_0=\sum_ih_{0i}$ and $H_1=\sum_ih_{1i}$, the maximum lies on the edge $\lambda_1=0$, with $\hat\lambda_0=T/H_0$, if every subject with $W_i>0$ has $h_{0i}>0$ and
\[
 \frac{H_0}{T}\sum_{i:\,W_i>0}\frac{W_ih_{1i}}{h_{0i}}\leq H_1 .
\]
Interchanging control and treatment gives the check for $\lambda_0=0$. Boundary datasets were identified with these checks; no dataset had zero events.
\begin{table}[!tbp]
\begin{singlespace}\centering\footnotesize\setlength{\tabcolsep}{2.3pt}
\caption{Count simulation: estimation of $\alpha_1=\log R$ under the Poisson count model in 72 settings varying the mean count per subject}\label{tab:sim_count_all72}
\begin{tabular}{@{}r@{\hspace{9pt}}r@{\hspace{6pt}}rrrrrr@{\hspace{10pt}}rrrrrr@{}}
\toprule
 & & \multicolumn{6}{c}{\boldmath$N=1{,}000$} & \multicolumn{6}{c}{\boldmath$N=100$} \\
\cmidrule(lr){3-8}\cmidrule(l){9-14}
\makecell[r]{$\boldsymbol{E(W_i)}$} & \makecell[r]{\textbf{Rate}\\\textbf{ratio}} & \makecell[r]{\textbf{Exp.\ trt}\\\textbf{events}} & \makecell[r]{\textbf{Boundary}\\\textbf{(\%)}} & \makecell[r]{\textbf{Valid}\\\textbf{fits}} & \textbf{Bias} & \textbf{RMSE} & \makecell[r]{\textbf{Cov.}\\\textbf{(\%)}} & \makecell[r]{\textbf{Exp.\ trt}\\\textbf{events}} & \makecell[r]{\textbf{Boundary}\\\textbf{(\%)}} & \makecell[r]{\textbf{Valid}\\\textbf{fits}} & \textbf{Bias} & \textbf{RMSE} & \makecell[r]{\textbf{Cov.}\\\textbf{(\%)}} \\
\midrule
\multicolumn{14}{@{}l}{\fontsize{8.5}{10}\selectfont\itshape{Short exposure, lower variability: $\mathrm{Beta}(1,30)$; mean 3.23\%, SD 3.12\%}} \\
\addlinespace[2pt]
0.3 & 0.5 & 4.17 & 39.38 & 6,062 & 0.823 & 1.395 & 76.2 & 0.42 & 49.12 & 5,078 & 2.042 & 2.397 & 63.8 \\
 & 1 & 8.23 & 30.50 & 6,950 & 0.323 & 1.109 & 84.7 & 0.82 & 45.08 & 5,487 & 1.390 & 1.863 & 73.8 \\
 & 2 & 16.01 & 14.63 & 8,537 & $-$0.045 & 0.965 & 91.4 & 1.60 & 40.02 & 5,989 & 0.846 & 1.511 & 81.0 \\
 & 7 & 49.45 & 0.13 & 9,987 & $-$0.066 & 0.483 & 95.8 & 4.94 & 18.86 & 8,108 & 0.073 & 1.145 & 93.6 \\
\addlinespace[2pt]
1 & 0.5 & 13.90 & 30.24 & 6,976 & 0.313 & 1.129 & 83.7 & 1.39 & 44.73 & 5,527 & 1.362 & 1.813 & 69.9 \\
 & 1 & 27.42 & 15.37 & 8,463 & $-$0.015 & 0.939 & 89.9 & 2.74 & 39.42 & 6,058 & 0.759 & 1.366 & 79.5 \\
 & 2 & 53.38 & 2.64 & 9,736 & $-$0.143 & 0.721 & 94.0 & 5.34 & 27.90 & 7,210 & 0.277 & 1.118 & 87.4 \\
 & 7 & 164.82 & 0.00 & 10,000 & $-$0.015 & 0.226 & 95.7 & 16.48 & 4.45 & 9,555 & $-$0.108 & 0.832 & 94.7 \\
\addlinespace[2pt]
3 & 0.5 & 41.70 & 18.78 & 8,122 & 0.005 & 0.996 & 87.9 & 4.17 & 39.55 & 6,045 & 0.831 & 1.408 & 76.2 \\
 & 1 & 82.26 & 4.37 & 9,563 & $-$0.125 & 0.761 & 93.3 & 8.23 & 29.28 & 7,072 & 0.330 & 1.116 & 84.9 \\
 & 2 & 160.13 & 0.05 & 9,995 & $-$0.055 & 0.385 & 94.8 & 16.01 & 15.23 & 8,477 & $-$0.022 & 0.954 & 90.9 \\
 & 7 & 494.46 & 0.00 & 10,000 & $-$0.006 & 0.128 & 95.1 & 49.45 & 0.25 & 9,975 & $-$0.071 & 0.492 & 95.7 \\
\addlinespace[5pt]
\multicolumn{14}{@{}l}{\fontsize{8.5}{10}\selectfont\itshape{Short exposure, higher variability: $\mathrm{Beta}(0.1,3)$; mean 3.23\%, SD 8.73\%}} \\
\addlinespace[2pt]
0.3 & 0.5 & 4.17 & 23.04 & 7,670 & 0.053 & 1.005 & 90.3 & 0.42 & 50.04 & 4,956 & 1.166 & 1.632 & 77.0 \\
 & 1 & 8.23 & 7.67 & 9,220 & $-$0.137 & 0.862 & 93.6 & 0.82 & 41.63 & 5,799 & 0.618 & 1.264 & 84.8 \\
 & 2 & 16.01 & 0.66 & 9,934 & $-$0.105 & 0.562 & 95.7 & 1.60 & 29.12 & 7,061 & 0.167 & 1.100 & 90.9 \\
 & 7 & 49.45 & 0.00 & 10,000 & $-$0.015 & 0.222 & 95.3 & 4.94 & 5.47 & 9,447 & $-$0.098 & 0.811 & 95.6 \\
\addlinespace[2pt]
1 & 0.5 & 13.90 & 6.62 & 9,332 & $-$0.130 & 0.799 & 93.6 & 1.39 & 36.12 & 6,385 & 0.477 & 1.234 & 83.8 \\
 & 1 & 27.42 & 0.40 & 9,958 & $-$0.090 & 0.516 & 95.1 & 2.74 & 23.43 & 7,653 & 0.099 & 1.019 & 90.2 \\
 & 2 & 53.38 & 0.00 & 10,000 & $-$0.025 & 0.256 & 95.8 & 5.34 & 10.41 & 8,959 & $-$0.089 & 0.875 & 93.6 \\
 & 7 & 164.82 & 0.00 & 10,000 & $-$0.005 & 0.116 & 95.5 & 16.48 & 0.25 & 9,975 & $-$0.068 & 0.483 & 96.1 \\
\addlinespace[2pt]
3 & 0.5 & 41.70 & 0.38 & 9,962 & $-$0.091 & 0.500 & 95.4 & 4.17 & 23.07 & 7,693 & 0.096 & 1.035 & 89.1 \\
 & 1 & 82.26 & 0.00 & 10,000 & $-$0.029 & 0.241 & 96.2 & 8.23 & 9.76 & 9,024 & $-$0.114 & 0.869 & 93.1 \\
 & 2 & 160.13 & 0.00 & 10,000 & $-$0.012 & 0.142 & 95.2 & 16.01 & 1.46 & 9,854 & $-$0.114 & 0.616 & 95.5 \\
 & 7 & 494.46 & 0.00 & 10,000 & $-$0.002 & 0.067 & 95.0 & 49.45 & 0.01 & 9,999 & $-$0.021 & 0.243 & 95.6 \\
\addlinespace[5pt]
\multicolumn{14}{@{}l}{\fontsize{8.5}{10}\selectfont\itshape{Longer exposure: $\mathrm{Beta}(2,5)$; mean 28.57\%, SD 15.97\%}} \\
\addlinespace[2pt]
0.3 & 0.5 & 41.46 & 0.81 & 9,919 & $-$0.105 & 0.601 & 95.7 & 4.15 & 23.45 & 7,624 & 0.214 & 1.201 & 91.7 \\
 & 1 & 72.86 & 0.00 & 10,000 & $-$0.036 & 0.353 & 96.4 & 7.29 & 11.40 & 8,844 & 0.001 & 1.066 & 96.1 \\
 & 2 & 117.24 & 0.00 & 10,000 & $-$0.008 & 0.261 & 95.3 & 11.72 & 3.41 & 9,655 & $-$0.029 & 0.961 & 98.6 \\
 & 7 & 207.56 & 0.00 & 10,000 & 0.015 & 0.256 & 95.3 & 20.76 & 8.04 & 9,186 & $-$0.002 & 0.872 & 99.5 \\
\addlinespace[2pt]
1 & 0.5 & 138.21 & 0.00 & 10,000 & $-$0.025 & 0.281 & 96.2 & 13.82 & 8.55 & 9,142 & $-$0.093 & 0.930 & 93.4 \\
 & 1 & 242.86 & 0.00 & 10,000 & $-$0.010 & 0.181 & 95.4 & 24.29 & 1.03 & 9,897 & $-$0.099 & 0.699 & 96.0 \\
 & 2 & 390.80 & 0.00 & 10,000 & $-$0.003 & 0.139 & 95.0 & 39.08 & 0.00 & 10,000 & $-$0.030 & 0.488 & 97.2 \\
 & 7 & 691.86 & 0.00 & 10,000 & 0.006 & 0.137 & 95.2 & 69.19 & 0.15 & 9,985 & 0.046 & 0.484 & 97.0 \\
\addlinespace[2pt]
3 & 0.5 & 414.63 & 0.00 & 10,000 & $-$0.009 & 0.153 & 95.6 & 41.46 & 0.85 & 9,915 & $-$0.114 & 0.648 & 95.3 \\
 & 1 & 728.57 & 0.00 & 10,000 & $-$0.004 & 0.101 & 95.5 & 72.86 & 0.00 & 10,000 & $-$0.037 & 0.357 & 96.5 \\
 & 2 & 1,172.41 & 0.00 & 10,000 & $-$0.002 & 0.079 & 95.5 & 117.24 & 0.00 & 10,000 & $-$0.013 & 0.265 & 95.5 \\
 & 7 & 2,075.58 & 0.00 & 10,000 & $-$0.001 & 0.079 & 94.7 & 207.56 & 0.00 & 10,000 & 0.017 & 0.257 & 95.4 \\
\bottomrule
\end{tabular}
\par\vspace{4pt}{\footnotesize\raggedright\setstretch{1.1}Each setting comprises 10,000 simulated datasets. Component counts are generated as $W_{0i}\sim\mathrm{Poisson}(\lambda_0h_{0i})$ and $W_{1i}\sim\mathrm{Poisson}(\lambda_1h_{1i})$, where $\lambda_0$ and $\lambda_1$ are the event rates under control and treatment, and only $W_i=W_{0i}+W_{1i}$ is analyzed. Here $h_{0i}+h_{1i}=1$ and $\Pr(h_{1i}>0)=0.85$; exposure means and SDs describe $h_{1i}$ conditional on $h_{1i}>0$. $E(W_i)$ is the mean count per subject and Exp.\ trt events is $N\lambda_1E(h_{1i})$; for a given exposure distribution, rate ratio and $E(W_i)$, the setting with $N=100$ has one tenth of the expected events of the setting with $N=1{,}000$. Boundary is the percentage of all 10,000 datasets whose likelihood is maximized at $\hat\lambda_1=0$ or $\hat\lambda_0=0$; these datasets were identified by the closed-form checks of Section~\ref{app:count-72}. No dataset had zero events. Estimates are obtained by direct Fisher scoring with step halving; valid fits are defined as in Section~\ref{sec:simulation_design} of the main paper. Bias, RMSE and Cov.\ (95\% Wald intervals with observed-information standard errors) summarize estimation of $\alpha_1=\log R$ and are computed over valid fits. The largest Monte Carlo standard errors across settings are approximately 0.018 for Bias, 0.50 percentage points for Boundary and 0.7 percentage points for Cov.\par}
\end{singlespace}
\end{table}

\FloatBarrier
\subsubsection{Quasi-Poisson analysis of overdispersed counts}\label{app:overdispersion}
Table~\ref{tab:count_overdispersion} gives results for the overdispersed counts of Section~\ref{sec:simulation_counts} of the main paper.
\begin{table}[!tbp]
\begin{singlespace}\centering\footnotesize\setlength{\tabcolsep}{3pt}
\caption{Count simulation under overdispersion: Poisson and quasi-Poisson inference for $\alpha_1=\log R$}\label{tab:count_overdispersion}
\begin{tabular}{@{}r@{\hspace{8pt}}lrrrrrrrr@{}}
\toprule
\makecell[r]{$\boldsymbol{E(W_i)}$} & \makecell[l]{\textbf{Mechanism}} & $\boldsymbol{\phi}$ & \makecell[r]{\textbf{Boundary}\\\textbf{(\%)}} & \textbf{Bias} & \textbf{RMSE} & \makecell[r]{\textbf{Cov.}\\\textbf{Poisson}} & \makecell[r]{\textbf{Cov.}\\\textbf{quasi}} & \makecell[r]{\textbf{Mean}\\$\boldsymbol{\hat\phi}$} & \makecell[r]{\textbf{Pearson}\\\textbf{rej.\ (\%)}} \\
\midrule
\multicolumn{10}{@{}l}{\fontsize{8.5}{10}\selectfont\itshape{Short exposure, lower variability: $\mathrm{Beta}(1,30)$; mean 3.23\%, SD 3.12\%}} \\
\addlinespace[2pt]
1 & Poisson & 1 & 2.64 & $-$0.143 & 0.721 & 94.0 & 94.0 & 1.00 & 5.1 \\
 & Component gamma & 1.5 & 5.79 & $-$0.131 & 0.832 & 89.5 & 93.1 & 1.50 & 100.0 \\
 & Component gamma & 3 & 12.96 & $-$0.075 & 0.953 & 80.0 & 91.1 & 2.99 & 100.0 \\
 & Subject frailty & 1.5 & 6.17 & $-$0.144 & 0.839 & 89.2 & 93.0 & 1.50 & 100.0 \\
 & Subject frailty & 3 & 13.81 & $-$0.054 & 0.946 & 78.5 & 90.9 & 2.99 & 100.0 \\
\addlinespace[2pt]
3 & Poisson & 1 & 0.05 & $-$0.055 & 0.385 & 94.8 & 94.8 & 1.00 & 5.0 \\
 & Component gamma & 1.5 & 0.36 & $-$0.084 & 0.493 & 91.6 & 95.0 & 1.50 & 100.0 \\
 & Component gamma & 3 & 2.53 & $-$0.148 & 0.757 & 83.1 & 93.8 & 3.00 & 100.0 \\
 & Subject frailty & 1.5 & 0.37 & $-$0.087 & 0.501 & 91.2 & 94.6 & 1.50 & 100.0 \\
 & Subject frailty & 3 & 2.76 & $-$0.128 & 0.723 & 82.2 & 93.4 & 3.00 & 100.0 \\
\addlinespace[5pt]
\multicolumn{10}{@{}l}{\fontsize{8.5}{10}\selectfont\itshape{Short exposure, higher variability: $\mathrm{Beta}(0.1,3)$; mean 3.23\%, SD 8.73\%}} \\
\addlinespace[2pt]
1 & Poisson & 1 & 0.00 & $-$0.025 & 0.256 & 95.8 & 95.7 & 1.00 & 5.2 \\
 & Component gamma & 1.5 & 0.00 & $-$0.045 & 0.330 & 91.0 & 95.9 & 1.50 & 100.0 \\
 & Component gamma & 3 & 0.16 & $-$0.098 & 0.525 & 81.1 & 95.2 & 2.99 & 100.0 \\
 & Subject frailty & 1.5 & 0.01 & $-$0.049 & 0.344 & 90.2 & 94.9 & 1.50 & 100.0 \\
 & Subject frailty & 3 & 0.50 & $-$0.104 & 0.557 & 79.4 & 94.2 & 2.99 & 100.0 \\
\addlinespace[2pt]
3 & Poisson & 1 & 0.00 & $-$0.012 & 0.142 & 95.2 & 95.2 & 1.00 & 5.3 \\
 & Component gamma & 1.5 & 0.00 & $-$0.013 & 0.176 & 89.3 & 95.4 & 1.50 & 100.0 \\
 & Component gamma & 3 & 0.00 & $-$0.028 & 0.254 & 75.2 & 95.9 & 2.99 & 100.0 \\
 & Subject frailty & 1.5 & 0.00 & $-$0.013 & 0.181 & 88.5 & 94.3 & 1.50 & 100.0 \\
 & Subject frailty & 3 & 0.00 & $-$0.035 & 0.284 & 70.5 & 94.0 & 2.99 & 100.0 \\
\addlinespace[5pt]
\multicolumn{10}{@{}l}{\fontsize{8.5}{10}\selectfont\itshape{Longer exposure: $\mathrm{Beta}(2,5)$; mean 28.57\%, SD 15.97\%}} \\
\addlinespace[2pt]
1 & Poisson & 1 & 0.00 & $-$0.003 & 0.139 & 95.0 & 94.8 & 1.00 & 5.1 \\
 & Component gamma & 1.5 & 0.00 & $-$0.003 & 0.172 & 89.0 & 95.0 & 1.50 & 100.0 \\
 & Component gamma & 3 & 0.00 & $-$0.005 & 0.242 & 75.5 & 95.5 & 2.99 & 100.0 \\
 & Subject frailty & 1.5 & 0.00 & $-$0.003 & 0.173 & 89.2 & 95.0 & 1.50 & 100.0 \\
 & Subject frailty & 3 & 0.00 & $-$0.008 & 0.245 & 74.6 & 95.3 & 2.98 & 100.0 \\
\addlinespace[2pt]
3 & Poisson & 1 & 0.00 & $-$0.002 & 0.079 & 95.5 & 95.5 & 1.00 & 5.0 \\
 & Component gamma & 1.5 & 0.00 & $-$0.001 & 0.099 & 88.6 & 95.0 & 1.50 & 100.0 \\
 & Component gamma & 3 & 0.00 & 0.001 & 0.138 & 74.5 & 95.0 & 3.00 & 100.0 \\
 & Subject frailty & 1.5 & 0.00 & $-$0.002 & 0.097 & 89.8 & 95.4 & 1.50 & 100.0 \\
 & Subject frailty & 3 & 0.00 & $-$0.003 & 0.140 & 74.2 & 94.7 & 3.00 & 100.0 \\
\bottomrule
\end{tabular}
\par\vspace{4pt}{\footnotesize\raggedright\setstretch{1.1}Settings have $N=1{,}000$ and $R=2$; each row is based on 10,000 simulated datasets, and rows with $\phi=1$ repeat the corresponding Poisson settings of Table~\ref{tab:sim_count_all72}. Component gamma: $W_{ki}\mid G_{ki}\sim\mathrm{Poisson}(G_{ki})$ with $G_{ki}\sim\mathrm{Gamma}$ of mean $\lambda_kh_{ki}$ and variance $(\phi-1)\lambda_kh_{ki}$, so that $\mathrm{Var}(W_i\mid h_i)=\phi\mu_i$, the quasi-Poisson assumption. Subject frailty: $W_{ki}\mid U_i\sim\mathrm{Poisson}(U_i\lambda_kh_{ki})$ with one $U_i\sim\mathrm{Gamma}$ of mean 1 and variance $1/k$ per subject, $k=E(W_i)/(\phi-1)$, so that $\mathrm{Var}(W_i\mid h_i)=\mu_i+\mu_i^2/k$ and the variance-to-mean ratio averages $\phi$ but differs between subjects. Subject frailty is a robustness check under a misspecified variance, for which a single scale factor need not restore nominal coverage. All fits use the Poisson likelihood. Cov.\ Poisson uses observed-information standard errors; Cov.\ quasi multiplies them by $\hat\phi^{1/2}$, with $\hat\phi=X^2/(N-2)$ and $X^2$ the Pearson statistic at the estimates; both use normal quantiles. Pearson rej.\ is the rejection rate of $X^2>\chi^2_{0.95,N-2}$. Boundary is over all datasets and was identified as in Table~\ref{tab:sim_count_all72}; the other columns are computed over valid fits. The largest Monte Carlo standard errors are approximately 0.010 for Bias and 0.5 percentage points for either coverage.\par}
\end{singlespace}
\end{table}

\FloatBarrier

\bibliographystyle{biometrika}
\bibliography{reference}